\documentclass[useAMS,usenatbib]{mn2e}

\usepackage{graphicx}
\usepackage{amssymb}
\newcommand{\eqb}{\begin{eqnarray}}
\newcommand{\eqe}{\end{eqnarray}}
\newcommand{\sth}{\sigma_{\rmn T}}

\newcommand{\diff}{{\rmn d}}

\newcommand{\spg}{\sigma_{\rmn p\gamma}^0}
\newcommand{\sgg}{\sigma_{\gamma \gamma}}

\newcommand{\tp}{\tau_{\rmn p}}
\newcommand{\tcr}{t_{\rmn {cr}}}
\newcommand{\Qpo}{Q_{\rmn {po}}}
\newcommand{\nex}{n_{\rmn {ex}}}
\newcommand{\nh}{n_{\rmn {h}}}
\newcommand{\np}{n_{\rmn {p}}}
 \newcommand{\ns}{n_{\rmn s}}
\newcommand{\gammamax}{\gamma_{\rm max}}
\newcommand{\gammamin}{\gamma_{\rm min}}

\newcommand{\gp}{\gamma_{\rmn p}}

\newcommand{\lhcr}{\ell_{\rmn h}^{\rmn cr}}

\newcommand{\eh}{\epsilon_{\rmn h}}
\newcommand{\es}{\epsilon_{\rmn s}}

\newcommand{\lB}{\ell_{\rmn B}}
\newcommand{\ls}{\ell_{\rmn s}}

\title[Feedback mechanisms and temporal behaviour of hadronic systems]
{Temporal signatures of leptohadronic feedback mechanisms in compact sources}
\author[M. Petropoulou and A. Mastichiadis]{M. Petropoulou  \thanks{E-mail:
maroulaaki@gmail.com } and A. Mastichiadis  \thanks{E-mail:
amastich@phys.uoa.gr}\\
Department of Physics, University of Athens, Panepistimiopolis, GR 15783 Zografos, Greece}
\begin{document}

\date{Received.../Accepted...}

\pagerange{\pageref{firstpage}--\pageref{lastpage}} \pubyear{...}

\maketitle

\label{firstpage}

\begin{abstract}
{The hadronic model of Active Galactic Nuclei and other compact high energy 
astrophysical sources assumes that ultra-relativistic protons, electron-positron pairs and photons
interact via various hadronic and electromagnetic processes inside 
a magnetized volume, producing the multiwavelength
spectra observed from these sources. 
A less studied property of such systems is that
they can exhibit a variety of temporal behaviours due to the operation of different
feedback mechanisms.}
{We investigate the effects of one possible feedback loop, where $\gamma$-rays produced
by photopion processes are being quenched whenever their compactness increases above a critical
level. This causes a spontaneous creation of soft photons in the system that result in further proton cooling and 
 more production of $\gamma$-rays, thus making the loop operate.
}
{We perform an analytical study of a simplified set of 
equations describing the system, in 
order to investigate the connection of 
its temporal behaviour with key physical parameters. 
We also perform numerical integration of the full set of kinetic equations 
verifying not only our analytical results but also those of previous numerical studies.}
{We find that once the system becomes `supercritical', 
it can exhibit either a periodic behaviour or a damped oscillatory one leading to
a steady state.  We briefly point out possible implications of such a supercriticality
on the parameter values used in Active Galactic Nuclei spectral modelling, through an indicative
fitting of the VHE emission of blazar 3C 279.}
\end{abstract}


\section{Introduction}

The idea that high energy protons can be produced in Active Galactic Nuclei (AGN) 
has been suggested by \cite{kazanas86} who considered 
proton acceleration by shock waves in the inner regions of these
objects. Following on this idea,
\cite{sikora87} suggested that, in this case, relativistic protons 
will lose energy in photohadronic interactions 
with the abundant soft photons 
rather than via inelastic collisions with the ambient 
cold protons. 
These ideas were later applied to the blazar jets
\citep{mannheim93, mueckeprotheroe01}
giving rise to what is known today as the hadronic model for
blazar high energy emission -- for a review see \cite{boettcher07} and \cite{boettcher10}.
Similar ideas have been applied also to Gamma Ray Bursts (\citealt{boettcherdermer98}, \citealt*{kazanasetal02}, \citealt{mastkazanas06, 
asanoinoue07}; \citealt*{asanoinoue09}; \citealt{mastkazanas09})
and $\gamma$-ray emitting compact binary 
systems\footnote{There is a difference between
hadronic models describing such systems and the corresponding ones used in modelling
 AGN or GRB high energy emission. 
The physical conditions in compact binary systems favour inelastic $pp$-collisions
 instead of photohadronic interactions, which for this reason are neglected.
}
(\citealt{romeroetal03}; \citealt*{paredesetal05, romeroetal05})
in order to explain the high energy emission from these objects. 
The hadronic models therefore form a viable alternative to the commonly used
leptonic ones.
 
High energy protons will radiate by synchrotron radiation, 
as well as by photopair and photopion while interacting on any soft photon present.
These interactions will produce secondaries: electron/positron pairs 
which are produced directly from photopair and via charged pion decay 
from photopion interactions as well as gamma-rays via neutral pion decay again 
from photopion; there will be also neutrino and 
neutron production coming as byproducts of photopion. While neutrinos 
will escape the source without any interactions and 
so their spectrum will be that at production, the created electron/positron 
pairs will lose energy through synchrotron radiation 
and inverse Compton scattering while gamma-rays will be absorbed in photon-photon collisions. 
Therefore in order to calculate the emerging 
photon spectrum one has to follow the evolution of these secondaries which can be
complicated due to the formation of
intense electromagnetic (EM) cascades initiated, e.g., by $\gamma-$rays from $\pi^0$ decay.

Usually the modelling of hadronic processes assumes that the target photons
come either from an external source or from synchrotron radiation of
a co-accelerated leptonic component.
One largely overlooked aspect is the possibility that protons interact with their 
own radiation, for example, with the soft photons produced from the aforementioned EM 
cascades. First attempts to incorporate these into the models were made by 
\cite{stern91} and \cite*{sternetal92}.
Using Monte Carlo simulations the above authors found that the system of protons 
and photons can exhibit limit cycles. 
However, this oscillating behaviour of the system could not be interpreted 
as the result of a specific feedback mechanism, 
let alone studied in a systematic way. One of the operating feedback processes was studied 
analytically by \cite{kirkmast92} using the kinetic equation approach\footnote{
A discussion about the different numerical approaches employed is presented in
 \cite{sternetal95}.}.
They showed that a sufficient number
density of protons can make the system unstable, causing runaway pair production: 
synchrotron photons of the relativistic electron-positron pairs
become targets for the protons which produce more pairs. The feedback 
leads eventually to fast proton energy losses. 
This amount of energy lost by the protons in a small
time interval is transfered to photons and it is seen as a flaring event.
The operation
of this feedback loop was later confirmed numerically by \cite{mastkirk95} and \cite*{mastkirk05}, who
also found cases where the system showed a limit cycle behaviour.

In the present paper we examine another type of feedback which can operate in
hadronic systems. For this we capitalize on the ideas of non-linear photon quenching \citep{SK07,PM11}.
According to this, $\gamma-$rays produced in a spherical volume cannot
exceed a critical luminosity that depends only on the source's magnetic
field and radius. If they do, then soft photons will be produced automatically
and quench the `excessive' $\gamma-$rays. 
In the context of a hadronic system
$\gamma-$rays can either be produced directly 
by proton synchrotron radiation or indirectly by photopair and photopion interactions. Then 
the following loop suggests itself:
\begin{enumerate}
\item Protons cool on soft photons producing $\gamma-$rays.\\
\item The $\gamma-$ray luminosity is quenched and turned spontaneously into 
soft photons which feedback on 1.
\end{enumerate}

Clearly this mechanism can tap energy stored in protons and transfer it
to radiation. At the same time it shows that the hadronic system is
a dynamical one and its behaviour can be more complex than it is
customarily assumed. A detailed study of the aforementioned feedback mechanism,
in the case where $\gamma$-rays are the byproduct of photopion interactions, 
will be the subject of the present work. The paper is structured as follows: 
In \S2 we describe qualitatively the system and define two regimes of operation. 
Next we construct a system of non-linear equations which we solve first analytically in a simplified
form (\S3). In \S4 we show in a semi-analytical way the role of various
processes on the dynamical behaviour of the system, while in \S5 we back
our results presenting a full numerical study of the problem. 
In \S6 we present an indicative astrophysical 
example, where some of our results are applied to the blazar 3C 279.
Finally, we conclude in \S7 with a summary and discussion.

\section{Qualitative description of the physical system}

\subsection{Linear regime} 
We assume a spherical source of radius R with embedded magnetic field B and a monoenergetic proton distribution
of number density $\tilde{n}_{\rmn p}$ and Lorentz factor $\gamma_{\rmn p}$.
This region is also filled with monoenergetic  radiation of energy $\epsilon_{\rmn o}$ 
(normalized to electron rest mass energy) 
 and number density $\tilde{n}_{\rmn {ex}}$\footnote{We note that here and through
 the present work tilted quantities
denote quantities with dimensions.}, 
which we will assume comes from outside of the source. Thus, we denote it as `external'.

The injected high energy protons will interact with the external photons through 
inelastic photopair and photopion collisions, provided that the respective threshold
conditions are satisfied. We will assume that the condition
\eqb
\epsilon_{\rmn o} \gp \gtrsim \frac{m_{\pi}}{m_{\rmn e}},
\label{eoth}
\eqe
where $m_{\pi}$ is the pion mass, is always satisfied.
This means that both photopair and photopion operate; however
since the target photons are monoenergetic, it guarantees
that photopion will be the main loss mechanism for protons (\citealt{sikora87}; \citealt*{begelman90}).
The produced charged and neutral pions will decay producing  electron/positron pairs 
(for brevity we will refer to them simply as `electrons') and $\gamma-$rays.
The former will radiate photons mainly through the synchrotron process, since inverse Compton scattering will
be greatly suppressed by Klein-Nishina effects.
 We will refer to these synchrotron photons
as `hard', since for proton energies with values typical of the AGN hadronic models, these can in principle
emerge in the $\gamma-$ray regime. 

One can quantify the above by noting that
the secondary electrons from the charged pion decay are produced with a Lorentz factor  
\eqb
\gamma_{\rmn e,\pi} \approx \eta_{\rmn p} \gp \frac{m_{\rm p}}{m_{\rmn e}},
\label{gepion}
\eqe
 where $\eta_{\rmn p}= k_{\rmn p}/4\simeq 0.08$ and $k_{\rmn p}$ 
is the inelasticity of the interaction assumed to be $\simeq 0.3$. 
The factor $1/4$  arises from the assumed energy
equipartition between the lepton and the 
three neutrinos produced by the charged pion decay -- see \cite{dimitrakoudisetal12}.
Assuming that these electrons emit at the critical synchrotron energy
\eqb
\eh=b\gamma^2_{\rmn e,\pi},
\label{eh}
\eqe
where $b=B/B_{\rmn cr}$ and $B_{\rmn cr}=4.413\times 10^{13}$ G the critical value of the magnetic field strength,
we find that for typical values of $\gamma_{\rmn p}=10^8$ and $B=1$~G, $\epsilon_{\rmn h}$ is in the TeV regime.

Let $\dot{\tilde{E}}_{\rmn tot}$ be the energy loss rate of all protons of energy $\gamma_{\rmn p}$ due to interactions with
the  photons. This energy is distributed to the produced secondaries. Assuming that their cooling is fast -- an assumption which is 
reasonable since both $B$ and $\gamma_{\rmn e}$ are assumed to have high values, one can argue that the
energy injected into secondary electrons will be instantaneously radiated as hard photons.
Then we can define the injected hard photon 
compactness as
\eqb
\ell_{\rm h}^{\rmn {inj}}=\xi_{\pi}{{\dot{ \tilde{E}}_{\rmn {tot}} \sth}\over{4 \pi R m_{\rmn e}c^3}},
\label{comph}
\eqe
where $\sth$ is the Thomson cross section and $\xi_{\pi}$ is the fraction of energy that goes to 
the secondary electrons. Furthermore, we can connect $\dot{\tilde{E}}_{\textrm {tot}}$ to the single
proton energy loss rate $\dot{\tilde{E}}_{\rmn p}$ through the relation
\eqb
\dot{\tilde{E}}_{\rmn{tot}}=\tilde{n}_{\rmn p} V \dot{\tilde{E}}_{\rmn p},
\label{totloss}
\eqe
where $V$ is the volume of the source.
Since photohadronic losses can be considered catastrophic, i.e., a relativistic proton can lose 
a substantial amount of its energy in one collision with a photon, we can write
\eqb
\dot {\tilde{E}}_{\rmn p} \simeq k_{\rmn p} \tilde{E}_{\rmn p} c \int \diff x \ \tilde{\sigma}_{\rmn p\gamma}(\gamma_{\rmn p} x) 
\tilde{n}_{\rmn ph}(x),
\label{indlossgen}
\eqe
where $\tilde{E}_{\rmn p}=\gp m_{\rmn p} c^2$ is the proton energy, $\tilde{\sigma}_{\rmn p\gamma}$ is the relevant cross section, 
$\tilde{n}_{\rmn ph}$ is the target photon population
and $x$ the target photon energy in units of $m_{\rmn e} c^2$. 
Under our assumptions, i.e., catastrophic energy proton losses
and monoenergetic particle distributions, we can adopt working, from this point on, with $k_{\rmn p}=1$
 without loss of generality. Furthermore, we can approximate the cross section  with a Heaviside function of the form
\eqb
\tilde{\sigma}_{\rmn p\gamma}(\gamma_{\rmn p},x) \simeq
\sigma_{\rmn p\gamma}^0 \sth H(\gamma_{\rmn p} x-m_{\pi}/m_{\rmn e})
\label{spg0}
\eqe
with
$\sigma_{\rmn p\gamma}^0=10^{-4}$; for a plot of the total expression of the cross
section see Fig.~3 in \cite{mueckeetal00}.
Using also the fact that
the only target photons present are the external ones, eq.(\ref{indlossgen}) becomes
\eqb
\dot{\tilde{E}}_{\rmn p}\simeq \gamma_{\rmn p} m_{\rmn p} c^2 
\sigma_{\rmn p\gamma}^0 c \tilde{n}_{\rmn{ex}}.
\label{indlossex}
\eqe
Combining relations (\ref{comph}) - (\ref{indlossex}) one can immediately deduce that the compactness
(or luminosity) of the hard photons depends on both $\tilde{n}_{\rmn p}$ and $\tilde{n}_{\rmn ex}$. 
In this case the system
can be considered to operate in the linear regime, since all cooling is provided by the external photons.
\subsection{Non-linear regime} 

The previous results indicate that for 
sufficiently high values of $\tilde{n}_{\rmn p}$ 
or $\tilde{n}_{\rmn {ex}}$, the injected hard photon compactness can take high values as well.
However, as it was shown in \cite{SK07} and \cite{PM11} -- henceforth SK07 and PM11 respectively, if the hard photon compactness is larger
than some critical value $\lhcr$ that depends only on $\eh$ and on source parameters such as  $B$ 
and $R$, 
even small initial perturbations of low energy photons present in the source, 
can grow and lead to an automatic quenching
of the hard photons. This is a purely non-linear process.
 In this case, electron-positron pairs 
grow spontaneously in the source and the `excessive'
hard radiation is absorbed by the synchrotron photons emitted by the pairs. Thus, a
soft photon population of number density $\tilde{n}_{\rmn s}$ and energy $\es$
appears spontaneously in the source. Assuming equipartition of energy between the created pairs one finds that
 $\es$ is
 given by
\eqb
\es = b \gamma_e^2=b \left(\frac{\eh}{2}\right)^2.
\label{es}
\eqe
These automatically produced photons have the same energy with those produced by the absorption of hard photons on the
external ones; note that this is a linear process.
Thus, both linear and non-linear absorption of hard photons result in the formation of a third photon population in the 
system with compactness $\ell_{\rmn s}$.  This new component will start playing a role in proton cooling through eq.~(\ref{indlossgen}),
since now $\tilde{n}_{\rmn {ph}}=\tilde{n}_{\rmn {ex}}+\tilde{n}_{\rmn s}$. 
If its number density grows sufficiently high, then it is possible
that the relativistic protons will start cooling more efficiently on them than on the external photons.
In this case, proton cooling becomes non-linear.

\begin{figure}
\centering 
\resizebox{\hsize}{!}{\includegraphics{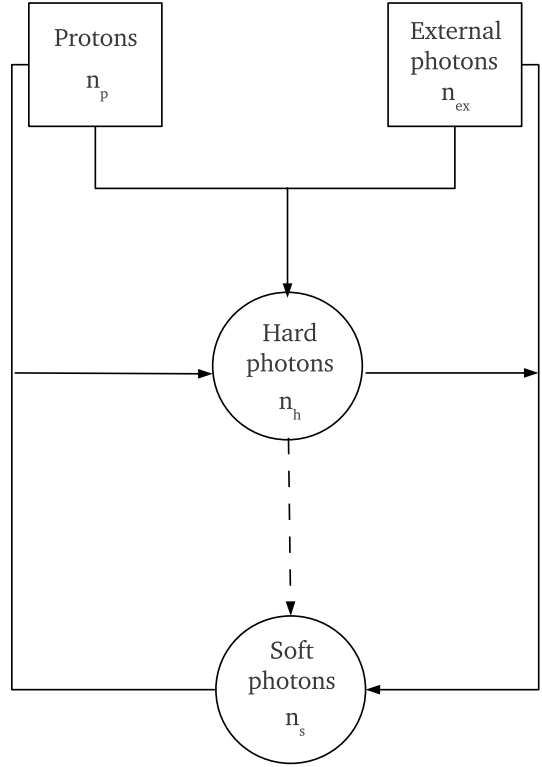}}
\caption{Schematic diagram of the operating loop of processes between protons and photons.}
\label{loop}
\end{figure}

Figure \ref{loop} summarizes the different processes operating in the system.
Arrows leading to the circles of Fig.~\ref{loop} denote 
injection of the corresponding particle species into the source, whereas arrows
coming out of the circle of hard photons imply their subsequent absorption.
In order to emphasize the existence of two absorbing channels for the $\gamma$-rays, the non-linear one
is shown with a dashed line. 
We note also that the secondary electrons that are the intermediate products of
the different processes operating in the system and
 responsible for the emission of hard and soft photons are not shown in Fig.~\ref{loop}.

We would like to examine next under which conditions the non-linear loop operates. 
There are two conditions
on the energies, that must be simultaneously satisfied. 
The first is the   
\textit{feedback criterion} for photon quenching. This
can be derived from the requirement that the magnetic field is
strong enough so that the synchrotron photons of the produced pairs are above the threshold for photon-photon absorption
on the hard photons. 
This leads to the condition for the magnetic field in the source $b>8\eh^{-3}$ (SK07; PM11). 
The last relation combined with eqs.~(\ref{gepion}), (\ref{eh}) and (\ref{es})
 sets a lower limit to the magnetic field strength which depends only on $\gp$
\eqb
b\gtrsim b_{\rmn q}=\left[\frac{\sqrt{8}}{\eta_{\rmn p}^3}\left(\frac{m_{\rmn e}}{m_{\rmn p}}\right)^3\right]^{1/2}\gp^{-3/2}.
\label{Blow1}
\eqe

The second is that the energy of the soft photons is high enough for the production of pions in interactions with the
protons, i.e., the relation 
\eqb
\es\gp \ge \frac{{m_\pi}}{m_{\rmn e}}
\label{crit2}
\eqe
should hold. 
This sets another lower limit to the magnetic field strength given by
\eqb
b \gtrsim b_{\pi}= \left(4 \frac{m_{\pi}}{m_{\rm e}}\right)^{1/3} \!\!\! 
\left(\eta_{\rmn p} \frac{m_{\rmn p}}{m_{\rmn e}} \right)^{-4/3}\gp^{-5/3}.
\label{Blow2}
\eqe

It is interesting to note that both $b_{\rmn q}$ and $b_{\pi}$ depend only on $\gamma_{\rmn p}$. 
Clearly, in order for non-linearity to appear in the system, the (normalized) magnetic
field of the source should satisfy the condition
\eqb
b\ge{\rm{max}}(b_{\rmn q},~b_{\pi}).
\label{Blowmax}
\eqe 
This is not a strict limit. For instance, if $\gamma_{\rmn p}=10^8$ one finds that 
$B_{\rmn q}=0.04$ G and $B_{\pi}=0.03$ G. Thus, magnetic fields 
of the order of 1 Gauss
can easily satisfy condition (\ref{Blowmax}).
Perhaps more limiting are other various effects, that we proceed to 
discuss next:
\begin{enumerate}
 \item[(i)]
 Photon quenching is based on the premise that soft photons start 
building in the system once the hard photons are above a certain critical luminosity, $\lhcr$.
The idea is that hard photons are absorbed on the automatically created soft photons; the produced
pairs emit more soft photons through synchrotron radiation which causes more
pair production on the hard photons etc. However, in the present situation,
the existence of an external photon population complicates the picture
as the hard photons might pair produce on them, in parallel to the internally
built soft photon population. This can act as a stabilizing factor, as it 
can inhibit the soft photon build-up in the system. This complication
can be avoided if one assumes that the relation $\eh\epsilon_{\rmn o}<2$ holds,
i.e., that collisions between the two photon populations are below the threshold
for pair-production.
 Since only $\eh$ depends on $B$ and $\gp$, one can find values for one of these
parameters, in order $\eh\epsilon_{\rmn o}<2$ to hold (see also $\S3$).
 In the analytical treatment of the next section
we will first start with this assumption but later we will relax this and study the
effects, that the hard photon pair production on the external ones
has on the dynamics of the system.\\ 
\item[(ii)]
 As the soft photons start building up, 
the secondary electron-positron pairs that produce them will start 
losing energy gradually through inverse Compton scattering instead of synchrotron
radiation. This means that the photon quenching loop could become
less efficient, because not all of the secondary electron luminosity will 
end up in the energy bin $\es$.
 This effect might become important near saturation, i.e., when the 
soft photon luminosity reaches its maximum value. As a first step, we will ignore
inverse Compton scattering in our analytical treatment. Then, we will
include it in an approximate manner to determine its effects.
Finally, we will take this fully into account in our
numerical treatment in $\S5$.\\
\item[(iii)]
 The exact nature of the external photon density in the source was of no importance so far. 
As a first order approximation to the problem, we neglect proton synchrotron radiation, assuming 
that the initial photon distribution is purely external. Under this assumption, both the external photon number density
 $n_{\rmn{ex}}$ and energy $\epsilon_{\rmn o}$
are treated as free parameters, instead of being determined by source properties, as the magnetic field B and the proton energy $\gp m_p c^2$.
However, in $\S5$, where  numerical solutions of the full problem are presented we 
replace the external source of photons by the proton synchrotron radiation.\\
\item[(iv)]
 In the analysis above the role of the hard photons was given to the synchrotron photons of
the charged pion decay. Without loss of generality we could also assign them to $\pi^0$ initiated secondaries:
Assuming that there is equipartition between the produced $\gamma-$rays from a $\pi^0-$decay and that
the inelasticity parameter is about the same as that for charged pion decay, we can
write in full analogy to eq.~(\ref{gepion}) 
\eqb
\epsilon_\gamma \approx 2 \eta_{\rmn p} \gp \frac{m_{\rmn p}}{m_{\rmn e}}.
\label{pi0}
\eqe
These are extremely hard $\gamma$-rays and will always be above the threshold
for pair production  on the external photons $\epsilon_{\rmn o}$.
The produced pairs will have energy $\gamma_{\rmn e,\pi^0}= \epsilon_\gamma / 2$ that is 
exactly the energy of the injected pairs through charged pion decay
(c.f. eq.~(\ref{gepion})). Therefore, both charged and neutral pions decay and
produce pairs of the same energy, that cool by synchrotron providing the 
hard photon emission. 
\end{enumerate}

\section{Analytical approach}
\subsection{Simplified equations}
In the most general case, the hadronic system consists of three species of particles, namely protons, electrons and photons. 
In order to
be described, the kinetic equations for the particles must be solved \citep{mastkirk95}.
Their generic form is:
\eqb
\frac{\partial\tilde{n}_i}{\partial t}+ \frac{\tilde{n}_i}{t_{i,\rmn{esc}}}=\tilde{L}^i+ \tilde{Q}^i,
\label{generic}
\eqe 
where the index $i$ can be one of the subscripts `p',`e' or `$\gamma$' referring to protons, electrons and photons respectively. 
The operators $\tilde{L}^i$ and $\tilde{Q}^i$ denote the losses and injection terms
respectively, whereas $\tilde{n}_i/t_{i,\rmn{esc}}$ is just
the escape term from the source, with each species having its own escape time $t_{i,\rmn{esc}}$. 
For photons the relation $t_{\gamma,\rmn{esc}}=t_{\rmn{cr}}=R/c$ holds, 
whereas for protons we adopt $t_{\rmn{p,esc}}=10^3 t_{\rmn{cr}}$ throughout the present work. 
The explicit expressions of the operators can be found
in \cite{mastkirk95} and \cite{mastkirk05} -- henceforth MK95 and MPK05 respectively.

 The unknown functions to be determined are the number densities $\tilde{n}_i$, that can be normalized as
follows:
\eqb
\tilde{n}_{\rmn p}(\tilde{E}_{\rmn p},\tau) & = & \frac{n_{\rmn p}(\gp,\tau)}{\sth R m_{\rmn p} c^2} \quad \textrm{with} \ \gp=\frac{\tilde{E}_{\rm p}}{m_{\rm p} c^2} \\
\phantom{1} \nonumber \\
\tilde{n}_{\rmn e}(\tilde{E}_{\rmn e},\tau) & = & \frac{n_{\rmn e}(\gamma,\tau)}{\sth R m_{\rmn e} c^2} \quad \textrm{with} \ \gamma=\frac{\tilde{E}_{\rm e}}{m_{\rm e} c^2} \\
\phantom{1} \nonumber \\
\tilde{n}_{\gamma}(\tilde{\epsilon}_{\gamma},\tau) & = & \frac{n_{\gamma}(\epsilon_{\gamma},\tau)}{\sth R m_{\rmn e} c^2} \quad
\textrm{with}\ \epsilon_{\gamma}=\frac{\tilde{\epsilon}_{\gamma}}{m_{\rmn e} c^2}.
\eqe
Time $\tau$ is normalized to the crossing time of the source $\tcr$, i.e., $\tau=t / \tcr$. From this point on, we adopt working with
dimensionless quantities.

\begin{figure} 
\resizebox{\hsize}{!}{\includegraphics{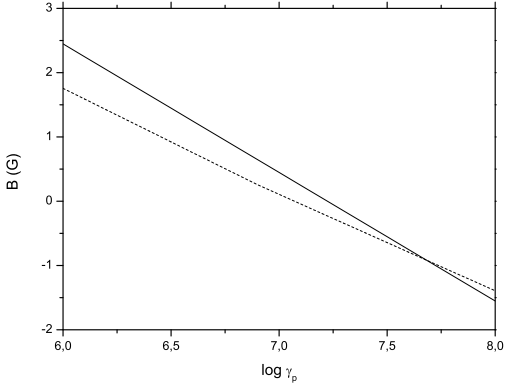}}
\caption{Upper (solid line) and lower (dashed line) limit of the magnetic field, where the lower limit is given
by $\max(B_{\rmn q},B_{\pi})$ (see eq.(\ref{Blowmax})). Any value of the magnetic field 
that lies below the solid and above the dashed line,
does not allow absorption
of hard photons by the external ones. 
The lines are drawn for the minimum energy of external photons that
satisfies the energy threshold for the photopion process, i.e. $\epsilon_{\rmn o}=\frac{m_{\pi}}{m_{\rmn e} \gp}$.}
\label{Bspace}
\end{figure}

For the purposes of an analytical treatment, we make two major simplifications:
\begin{enumerate} 
\item[(i)]
Since electron cooling can be considered fast for typical values of the system's parameters
 (see also $\S2$),  we can neglect the equation of the electrons. \\
\item[(ii)]
We simplify the equations describing the physical system to such a point
as to retain only the key processes.  
\end{enumerate}
Thus, we assume a constant monoenergetic injection
of protons $\Qpo(\gamma_{\rmn p})$ into a spherical source of radius R and embedded magnetic field B. 
We assume also a monoenergetic distribution of external photons $\nex$ at energy $\epsilon_{\rmn o}=\gp^{-1}\frac{m_{\pi}}{m_{\rmn e}}$,
 that acts as a target for high energy protons.
We note that initially, $\nex$ is the only distribution of low energy photons present in the source.
Charged pions produced by proton-photon pion processes, decay into electron-positron pairs with Lorentz factor $\gamma_{\rmn e,\pi}$ 
(see eq.~(\ref{gepion})), that emit through
synchrotron radiation, hard photons with corresponding number density $\nh(\eh)$. 
In order to ensure that inverse Compton scattering of the external photons by the aforementioned electrons is not as important
as synchrotron cooling, and therefore can be safely neglected,
we assume that $u_{\rmn {ex}}<<u_{\rmn B} $ or equivalently $\ell_{\rmn {ex}} << \ell_{\rmn B}$, where $u_{i}$ and $\ell_{i}$ are the energy densities and compactnesses
 respectively.
At this point, it is useful to define the different compactnesses that appear in the present work:
\eqb
\ell_{i}& = &\frac{\epsilon_{i} n_{i}}{3} , \quad i = \textrm{s, h, ex} \\
\ell_p & = & \frac{\gp \np}{3}\\
\ell_B & = & \sth R \frac{u_{\rmn B}}{m_{\rmn e} c^2}.
\eqe
 In our treatment, secondary electrons do not affect the dynamics of the system. They rather 
play an intermediate role for transferring energy
from hard photons to lower energy photons. 
 We examine separately the hard and soft photon populations, $\nh$ and $\ns$ respectively,
 by writing a kinetic equation for each one of them. Thus, the quantities
to be determined now, are $\np$, $\nh$ and $\ns$.
The physical processes to be included into the simplified version of equations are:
\begin{enumerate}
\item Constant proton injection $\Qpo$ and proton escape $ L^{\rmn p}_{\rmn {esc}}= -\np / \tau_{\rmn p}$, 
that act as a source and a loss term respectively in the proton equation.
 \item Proton-photon pion production, that acts as a loss term for protons $L^{\rmn p}_{\rmn p \gamma \rightarrow \rmn p \pi}$ 
and as an injection term for hard photons $Q^{\rmn h}_{\rmn p \gamma \rightarrow \rmn p \pi}$.
\item Photon-photon pair production, that acts as an absorption term for hard photons $L^{\rmn h}_{\gamma \gamma}$ and as an
injection term for soft photons $Q^{\rmn s}_{\gamma \gamma}$.
\item Photon escape from the source in a crossing time, i.e., $ L^{\gamma}_{\rmn {esc}}= -\np$.
\end{enumerate}
Under these considerations, the simplified kinetic equations for each species are given by:
\eqb
\dot{n}_{\rmn p} & = & \Qpo -\frac{\np}{\tp} + L^{\rmn p}_{\rmn p \gamma \rightarrow \rmn p \pi} 
\label{np0}
\eqe
\eqb
\dot{n}_{\rmn h} & = & -\nh + Q^{\rmn h}_{\rmn p \gamma \rightarrow \rmn p \pi}+ L^{\rmn h}_{\gamma \gamma} 
\label{nh0}
\eqe
\eqb
\dot{n}_{\rmn s} & = & -\ns + Q^{\rmn s}_{\gamma \gamma}.
\label{ns0}
\eqe
The analytical study of the dynamics of the above set of equations can be simplified even further, if we assume that 
hard photons cannot be absorbed by the external ones (see point (i) of $\S2$). 
To ensure this, we use for a given $\gp$, such values of B
that do not allow further absorption of hard photons. Thus, the condition $\eh \epsilon_{\rmn o} <2$ should hold. 
This, combined with eq.~(\ref{gepion}) and 
(\ref{eh}) sets an upper limit for the magnetic field strength
\eqb
b\lesssim b_{\alpha}= 2\frac{m_{\rmn e}}{m_{\pi}}\left(\frac{m_{\rmn e}}{\eta_{\rmn p} m_{\rmn p}}\right)^2.
\label{Bup}
\eqe
Equations (\ref{Blow1}), (\ref{Blow2}) and (\ref{Bup}) can be combined in order to create a parameter space 
of allowed values of B for different
proton energies -- see Fig.~\ref{Bspace}. Solid line shows the upper limit of eq.(\ref{Bup}), whereas dashed line
shows the maximum value of the two lower limits given by eq.~(\ref{Blowmax}). 

According to $\S2$, if the compactness of hard photons is below the critical value $\lhcr$, the instability that leads to an `automatic' quenching
of hard photons cannot grow. 
Thus, there is no injection of soft photons in the system and hard photons do not
suffer any absorption. This situation corresponds to the linear regime described in $\S 2.1$.
In this case, assuming that
proton losses
are catastrophic we can write
 $ L^{\rmn p}_{\rmn p \gamma \rightarrow \rmn p \pi} = -\spg \np \nex$ and $Q^{\rmn h}_{\rmn p \gamma \rightarrow \rmn p \pi}=A \np \nex$.
The set of equations 
(\ref{np0})-(\ref{ns0}) degenerates into a system of two equations (S1):
 \eqb
\dot{n}_{\rmn p} & = & \Qpo -\frac{\np}{\tp}-\spg \np \nex
\label{np1}
\eqe
\eqb
\dot{n}_{\rmn h} & = & -\nh + A \np \nex.
\label{nh1}
\eqe
The normalization constant $A$ of the injection term in the hard photon equation above, is found after assuming
 that a fraction $\xi_{\pi}$ of the total proton energy goes to secondary electrons, that 
radiate further all their energy through synchrotron producing the hard photons. Under these considerations, one finds
that $A=\xi_{\pi}  \frac{\spg m_{\rmn p} \gp}{m_{\rmn e} \eh}$.
The steady state solution of system S1 is
\eqb
\np^{\rmn {ss}} & = & \frac{\Qpo}{G_{\rmn p}} 
\label{npss}
\eqe
\eqb
\nh^{\rmn {ss}} & = & A \nex \np^{\rmn {ss}},
\label{nhss}
\eqe
where $G_{\rmn p}=\frac{1}{\tp}+\spg \nex$. 
Two limiting cases can be implied by the form of $G_{\rmn p}$:
\begin{itemize} 
\item[\bf{.}] 
Proton escape is more significant than proton cooling on the external photons, i.e., $\spg \nex \ll 1/\tp$.
In this limit, the steady state solution for hard photons is proportional to the product $\Qpo \nex$.
One can find a combination of values for $\Qpo$ and $\nex$ that lead to $\ell_{\rmn h} > \lhcr$. For example if proton cooling
is not efficient in injecting hard photons into the system because of a low $\nex$, a high value of the proton injection rate is needed
and vice versa.
\\
\item[\bf{.}]
Proton cooling on the external photons is more significant than proton escape, i.e., $\spg \nex \gtrsim 1/\tp$.
In this limit, $\nh^{\rmn {ss}} \propto \Qpo$. Thus, the system can become supercritical for a high enough
value of the proton injection rate.
\end{itemize}

So far, the proton-photon system operates in a linear `subcritical' regime with a very
well described behaviour. In order to investigate how the non-linear terms affect the
evolution of the system, we will focus only on cases where quenching is relevant.
If hard photons are being injected into the
system with $\ell_{\rmn h}^{\rmn {inj}} > \lhcr$, 
then a soft photon population $\ns(\es)$ appears because of the quenching of the $\gamma$-rays and the
system becomes `supercritical'.
 In this case an additional equation for the soft photons is required. 
Thus, the set of equations (\ref{np1})-(\ref{nh1}) becomes (system S2):
\eqb
\dot{n}_{\rmn p} & = & \Qpo -\frac{\np}{\tp}-\spg \np \nex - \spg \np \ns
\label{proton1}
\eqe
\eqb
\dot{n}_{\rmn h} & = & -\nh + A \np \nex + A \np \ns - C_{\rmn h} \ns \nh 
\label{hard1}
\eqe
\eqb
\dot{n}_{\rmn s} & = & -\ns +C_{\rmn s} \ns \nh,
\label{soft1}
\eqe
where
\eqb
C_{\rmn h}=\frac{\sgg^{(\rmn s)}}{\eh \es} \qquad \textrm{and} \qquad C_{\rmn s}=\frac{\sgg^{(\rmn s)}}{\es^2}.
\label{constants1}\eqe
The term $\sgg^{(s)}$ that appears in expressions (\ref{constants1}) is
given by
\eqb
\sgg^{(\rmn s)}= \eh\es \sgg(\eh \es)
\eqe
with the exponent `s' denoting the absorption on the soft photons and $\sgg$ 
 the cross section of photon-photon absorption, measured in units of the Thomson
cross section $\sigma_T$.
Here and throughout this work, we use the approximate expression of \cite{coppibland90}:
\eqb
\sgg(x)= 0.652 \frac{(x^2-1)}{x^3} \ln(x) H(x-1),
\label{sgg}
\eqe
where $x$ is the product of the dimensionless photon energies and $H(x)$ is the Heaviside step function. 
Since in our analysis we have assumed monoenergetic photon distributions and a cross section approximated by a step function,
$\sgg^{(\rmn s)}$ in definitions (\ref{constants1}) is just a constant, that takes different values for different pairs of photons. 
 The last terms in the right hand side of eqs. (\ref{hard1}) and (\ref{soft1})
 account for the photon-photon absorption. The created pairs are the intermediate products that
 transport their energy through synchrotron radiation to soft photons. 
The constants $C_{\rmn h}$ and $C_{\rmn s}$
given above are determined by energy conservation.
As far as the soft photons produced have
enough energy to satisfy the energy threshold for photopion interactions, 
(see relation (\ref{crit2})), they can act as an additional target for protons. 
This explains the last term in the right hand side of eq.~(\ref{proton1}). This additional sink term
of protons has a corresponding injection term  $A \np \ns$,
 that appears in eq.~(\ref{hard1}).

We note that an expression of the critical compactness for the automatic quenching of hard photons has been presented
by SK07 and PM11. In the present analysis we have made certain simplifying assumptions that differ from those in the aforementioned
papers. Thus, for reasons of consistency, we calculate the modified expression of $\lhcr$ 
\eqb
\lhcr=\frac{\eh}{3 C_{\rmn s}}
\eqe
(see Appendix C for the derivation).

\subsection{Dynamical study of the system}
After having set the framework of the physical problem and clarified our assumptions, we proceed to 
investigate the mathematical properties of the system.

Setting in equations (\ref{proton1})-(\ref{soft1}) the time derivatives equal to zero we can determine
the possible steady states, that actually are the `fixed points' of
the dynamical system.
The system of equations S2 has one or three fixed points depending on the existence or not
of the coupling terms between hard and soft photons. 
We will treat the system of equations that obtains three fixed points, since this can in principle show 
non-linear temporal behaviour.

\begin{figure}
\resizebox{\hsize}{!}{\includegraphics{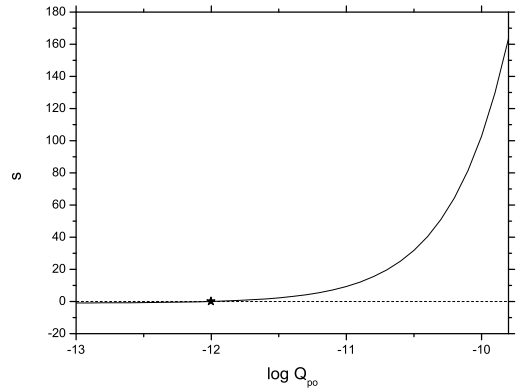}}
 \caption{Growth/decay rate of the perturbed proton and photon densities from the 
steady state $P1$ as a function of the proton injection rate. Above a certain value of $\Qpo$, that is denoted with a star,
only growing perturbations exist.
Parameters
used for this plot are: $B=0.7$ G, $\gp=2 \times 10^7$, $\epsilon_{\rmn o}=\gp^{-1} \frac{m_{\pi}}{m_{\rmn e}}$ and $\nex=2$.}
\label{fig1}
\end{figure}

The first fixed point $P_1$ is just the steady state solution presented in the previous section, 
i.e., $P_1 \left(\nh^{\rmn {ss}},0,\np^{\rmn {ss}}\right)$. Perturbations of this steady state solution can either grow or decay with time as $e^{\rmn s\tau}$.
In the first case, the fixed point $P_1$ is unstable whereas in the second case it is stable (see Appendix A for a detailed analysis).
The growth or decay rate $s$ of the perturbations as a function of the proton injection rate,
 for a fixed external number density, is shown in Fig.~\ref{fig1}. 
 It is interesting to note that the exponent $s$ 
  for the growing solutions depends strongly on $\Qpo$, whereas the respective one for the decaying solutions is rather insensitive to $\Qpo$ and 
close to $-1$, denoting the free escape of the produced soft photons in a crossing time. 
 The star denotes the critical value of the proton injection rate, $\Qpo^{\rmn {cr}}$, above which $s>0$ and an automatic build-up
of soft photons in the system is possible.
This value depends on the external photon density as:

\eqb
\Qpo^{\rmn {cr}}(\nex) = \frac{1}{C_{\rmn s} A \nex}\left(\frac{1}{\tp}+\spg \nex \right).
\label{Qpmin}
\eqe
Low values of $\nex$ suggest inefficient proton cooling and therefore inefficient 
injection of hard photons -- see eqs.~(\ref{np1})-(\ref{nh1}). Only if the injection rate of  protons in the source
is high, can the hard photon compactness increase sufficiently leading the system to instability. 
In other words, for a given $\nex$ one can find always a sufficiently high $\Qpo$ in order to make the system unstable.
$\Qpo^{\rmn {cr}}$ is
directly related to a critical proton number density $\np^{\rmn {cr}}$ given by
\eqb
\np^{\rmn {cr}} = \frac{\Qpo^{\rmn {cr}}}{1/\tp +\spg \nex}.
\eqe
Thus, the problem of choosing a suitable pair of values $(\Qpo,\nex)$ that lead the system to instability, can be transferred
to the problem of loading the source with a critical proton content.
Let us now consider the inverse of the above statement: \textit{if the
injection rate of protons is given, is there any value of $\nex$ that will make the system unstable?}
In this case, we find that the external number density
must satisfy the following condition:
\eqb
\nex \gtrsim \frac{1/ \tp}{C_{\rmn s} A \Qpo -\spg},
\eqe
with the constraint $\Qpo > \frac{\spg}{C_{\rmn s} A}$. 
Thus, for low enough proton injection rates, the system cannot become unstable. The above remarks lead to the conclusion
that, between the two parameters, $\Qpo$ and $\nex$, the fundamental one in making the system unstable is the proton
content of the source.

When the initial growth of the perturbations is ensured, the subsequent behaviour of the system depends on the properties of the second
fixed point $P_2$. The system can either reach the steady state $P_2$ or vary periodically, making a limit cycle in phase space.
This behaviour can be predicted by calculating the eigenvalues of the matrix $M_2$ of the linearized system of equations,
near the point $P_2$. For different values of the physical parameters of the problem, we find always one
real negative eigenvalue, $\lambda_1$. The two remaining eigenvalues $\lambda_{2,3}$ can either be complex 
conjugates or both real and negative (see Appendix B for more details).
Table~1 summarizes the dynamical behaviour of the system. We note that, 
for the terminology regarding the classification of a fixed point of a three-dimensional system,
we have adopted the one presented in \cite*{theisel03}.

\begin{center}
\scalebox{0.75}{%
 \small
\begin{tabular}{c c c}
\hline 
\multicolumn{3}{c}{\phantom{1}} \\
\textrm{\large{Eigenvalues}} & \textrm{\large{Classification of point $P_2$}} & \textrm{\large{Dynamical behaviour}} \\
\hline
\multicolumn{3}{c}{\phantom{1}} \\ 
$\lambda_i <0$, \textrm{for} $i=1,3$ & \textrm{\normalsize{attracting}} \textit{\normalsize{node}} & \textrm{\normalsize{steady state}} \\
\multicolumn{3}{c}{\phantom{1}} \\
\hline
\hline
\multicolumn{3}{c}{\phantom{1}} \\
$ \lambda_1 <0$, $\lambda_2=\lambda_3^{*}$ & \textit{\normalsize{focus}} & \phantom{1}\\
\multicolumn{3}{c}{\phantom{1}} \\
\hline 
\multicolumn{1}{c|}{\phantom{$\textrm{Re}(\lambda_2)=\textrm{Re}(\lambda_3)>0$}} & \multicolumn{1}{|c|}{\phantom{repelling saddle}} & \multicolumn{1}{|c}{\phantom{limit cycles}} \\
\multicolumn{1}{c|}{$\textrm{Re}(\lambda_2)=\textrm{Re}(\lambda_3)>0$} & \multicolumn{1}{|c|}{\textrm{\normalsize{repelling saddle}}} & \multicolumn{1}{|c}{\textrm {\normalsize{limit cycles}}}\\
\multicolumn{1}{c|}{\phantom{$\textrm{Re}(\lambda_2)=\textrm{Re}(\lambda_3)>0$}} & \multicolumn{1}{|c|}{\phantom{repelling saddle}} & \multicolumn{1}{|c}{\phantom{limit cycles}} \\
\multicolumn{1}{c|}{$\textrm{Re}(\lambda_2)=\textrm{Re}(\lambda_3)<0$} & \multicolumn{1}{|c|}{\textrm{\normalsize{attracting}}} & \multicolumn{1}{|c}{\textrm {\normalsize{damped oscillations}}}\\
\hline
\end{tabular}}
\end{center}
\footnotesize{\textbf{Table 1}: Classification of the fixed point $P_2$ and of the expected dynamical behaviour of the system, 
based on an eigenvalue/eigenvector
analysis of the corresponding matrix $M_2$.}
\vspace{0.4cm}

\normalsize
It is interesting to investigate how the two main parameters of the physical problem, i.e., the proton injection rate $\Qpo$ and the 
external number density $\nex$, are related to the dynamical behaviour of the system. 
For this purpose, we calculate the eigenvalues of matrix $M_2$ around the fixed point $P_2$ for different values of $\Qpo$ and $\nex$, having first ensured that the
fixed point of the steady state $P_1$ is unstable (see Appendix A). 
The results are presented in the next two paragraphs.
\subsubsection{Dependence on proton injection rate}
\begin{figure}
\resizebox{\hsize}{!}{\includegraphics{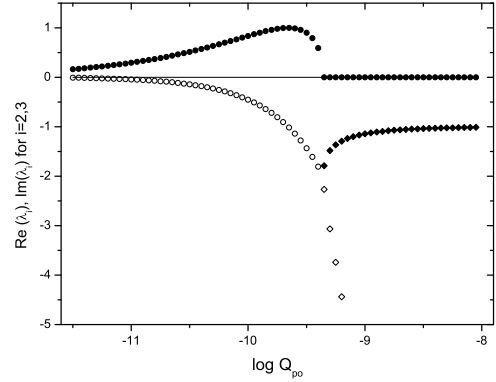}}
\caption{Plot of the two eigenvalues $\lambda_{2,3}$ for $\nex=2$ as a function of the injected proton rate. 
As long as the eigenvalues are complex conjugates, their real and imaginary parts are shown with open and filled circles respectively.
At a certain value of $\Qpo$ both eigenvalues become real and negative (open and filled diamonds). The solid line corresponds to 
the null value. The other parameters used are the same as in Fig.~\ref{fig1}.}
\label{fig2}
\end{figure}

\begin{figure}
\resizebox{\hsize}{!}{\includegraphics{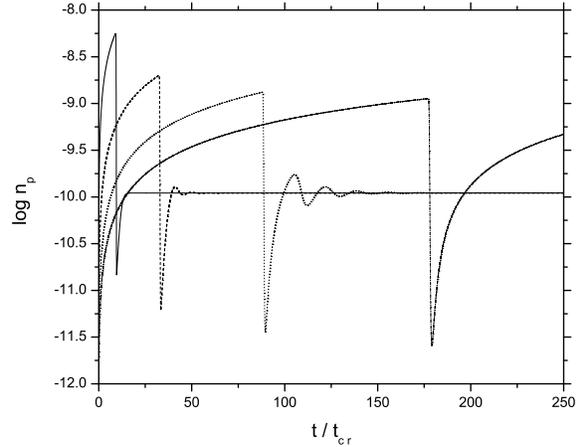}}
\caption{Time evolution of the proton distribution for different values of the proton 
injection rate $\log \Qpo= -11.15$ (dashed-dotted line), 
$\log \Qpo= -10.8$ (dotted line), $\log \Qpo=-10.2$ (dashed line) and $\log \Qpo=-9.2$ (solid line). 
The initial conditions for each numerical run are $\nh(0)=\np(0)=0$ and $\ns(0)=\epsilon \rightarrow 0$. 
The number density of the external photon distribution is $\nex=2$ in all cases. 
All the other parameters used are the same as in Fig.~\ref{fig1}. 
}
\label{figclose}
\end{figure}

\begin{figure}
\resizebox{\hsize}{!}{\includegraphics{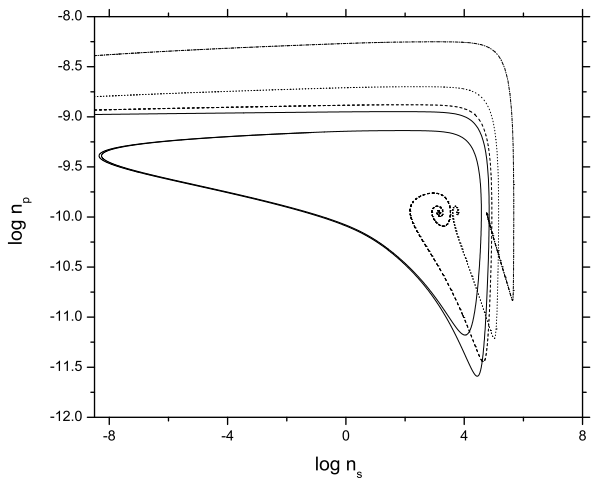}}
\caption{Two-dimensional plane of the phase space for different values of the proton injection rate $\log \Qpo= -11.15$ (dashed-dotted line), 
$\log \Qpo= -10.8$ (dashed line), $\log \Qpo=-10.2$ (dotted line) and $\log \Qpo=-9.2$ (solid line). Same parameters used
as in Fig.~\ref{figclose}.}
\label{figphclose}
\end{figure}

We assume first that the external density $ \nex$ is fixed to a certain value and study
the effects of the injection rate $\Qpo$. 
Figure \ref{fig2} shows the calculated eigenvalues $\lambda_{2,3}$ for different values of $\Qpo$. 
The conclusions drawn from it can be summarized
in the following points:
\begin{enumerate}
 \item
Starting with low values of $\Qpo$, we find that $\lambda_3=\lambda_2^{*}$. 
As long as $\textrm{Re($\lambda$)} > 0$,
 point $P_2$ acts
as a repelling focus - saddle point. Having also ensured that the fixed point of the steady state $P_1$ is unstable,
the system follows in phase space a closed periodic trajectory, i.e., a limit cycle that is moreover stable.  \\
\item
For higher values 
of the proton injection rate we find complex eigenvalues with $\textrm{Re($\lambda)$}<0$. 
Therefore, the point $P_2$ acts as an attracting focus - saddle
point. In this case, the 
system settles down in a new steady state (given by the fixed point $P_2$). In phase space, this
 corresponds to a `spiraling' trajectory ending at the fixed 
point.\\
\item
 Finally, for even higher values of $\Qpo$ both eigenvalues become real and negative. 
Point $P_2$ can be characterized
as an attracting node. The physical system reaches the new steady state very fast, showing no oscillations.
\end{enumerate}
An example of such a transition in the dynamics of the system,
 is shown in figures \ref{figclose} and \ref{figphclose}. 
The solution shown with dashed-dotted line corresponds to a limit cycle case with period  $\sim 170 \ \tcr$. 
For reasons of clarity, Fig.~\ref{figclose} zooms in the early time behaviour of the system. Therefore, 
only the first and a half cycle of the periodic solution is shown.
The solutions presented in the aforementioned
figures are found after integrating the system of equations S2 with initial conditions $\nh(0)=\np(0)=0$ and 
$\ns(0)=\epsilon$, with 
$\epsilon \rightarrow 0$ indicating an initial perturbation of soft photons in the source.
 The parameters used ensure that at some point the compactness of hard photons becomes larger than $\lhcr$.
Each of the values of $\Qpo$ used in this example corresponds
to a different point of Fig.~\ref{fig2}. 
The conclusion is that the temporal behaviour of the system is very sensitive with respect to the proton injection rate.

For reasons of mathematical consistency we make also the following remark: 
As the proton injection rate increases, the real part of the complex eigenvalues changes sign and from positive it turns
into negative. At this transition 
 a second unstable limit cycle\footnote{We note that the existence of the 
unstable limit cycle was found numerically while integrating the equations of system S2 and not analytically.} 
 surrounding the fixed point $P_2$ appears inside the stable limit cycle (see point (1) above). 
An integration of the equations of S2 with initial conditions lying outside the stable limit cycle would
show that the system falls onto this cycle instead of `spiraling' down to the fixed point, as one would expect 
according 
to point (2).
As the varying parameter $\Qpo$ increases further, $|\textrm{Re}(\lambda_{2,3})|$ increases too and
the unstable limit cycle approaches the stable one. At some value, which for the specific example of Fig.~\ref{fig2} is 
$\log \Qpo=-10.9$, the two 
limit cycles coalesce and the system undergoes a bifurcation\footnote{The existence of the bifurcation is not
a characteristic property of the system for all values of the parameters. For example, if $\nex=10$ no bifurcation
of this type is found.}
since the phase space changes qualitatively. From this moment on the system can
fall into the fixed point $P_2$, as described in (2) above.
Thus, the condition $\textrm{Re}(\lambda_{2,3})<0$ that occurs for
$\log \Qpo = -11.5$ in our example, does not ensure strictly speaking the damped oscillatory behaviour of 
the system. 
 
It is beyond the scope of the present work to proceed into a detailed study of the bifurcation
mentioned earlier, since for the initial conditions of physical relevance to our analysis, the existence
of the bifurcation does not affect the qualitative features of the transition from the limit cycle phase to the 
damped oscillatory one.

\subsubsection{Dependence on number density of external photons}
\begin{figure}
\resizebox{\hsize}{!}{\includegraphics{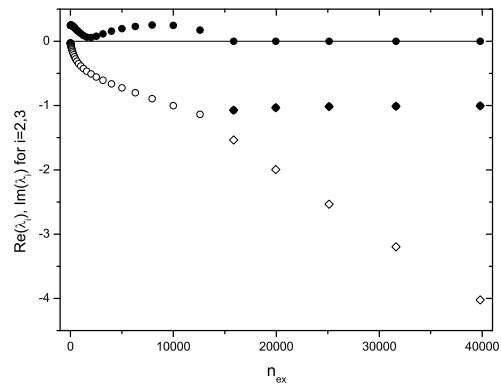}}
\caption{Plot of the two eigenvalues $\lambda_{2,3}$ as a function of the external number density $\nex$ for injected proton rate 
$\log \Qpo = -11.15$. 
While the eigenvalues are complex conjugates, their real and imaginary parts are shown with open and filled circles respectively.
At a certain value of $\nex$ both eigenvalues become real and negative (open and filled diamonds). The solid line corresponds to 
the null value. All other parameters used are the same as in Fig.~\ref{fig1}.}
\label{fig3}
\end{figure}

In this section we show an example of the effects of the external number density $\nex$ on the dynamical properties of the system. 
An increase of 
the external number density 
leads to the same transition of the dynamical behaviour of the system, as the one described in the previous section.
 Figure 
\ref{fig3} shows the calculated eigenvalues for a large range of $\nex$ extending up to high values. The qualitative features of this plot
are the same as those of Fig.~\ref{fig2}. 
One should keep in mind though, that the conclusions drawn from Fig.~\ref{fig3},
 regarding the dynamics of the system are not valid for the whole range of values of $\nex$ shown. The reason is the following:
As $\nex$ increases, the compactness
of the external photons $\ell_{\rmn {ex}}$ increases too. At some value, which for the specific example is just $\nex= 35$, 
it becomes larger than the compactness
of the magnetic field $\lB=1.6\times 10^{-4}$. 
This means that the secondary electrons produced by the pion decay, cool preferably through
inverse Compton scattering on the external photon field, rather than through synchrotron radiation. 
Thus, the system of equations S2 we used
to make our mathematical analysis is physically not valid. However, aim of Fig.~\ref{fig3} is to simply show the 
mathematical similiraties of this case with the previous one.

\begin{figure}
\resizebox{\hsize}{!}{\includegraphics{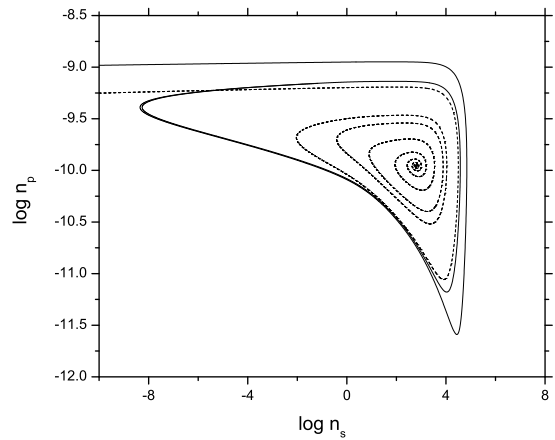}}
\caption{Two dimensional plane of the phase space for two values of the external number density $\nex=2$ (solid line) and 
$\nex=4$ (dashed line). The proton injection rate is $\log \Qpo = -11.15$ in both cases.
The initial conditions for each numerical run are $\nh(0)=\np(0)=0$ and $\ns(0)=\epsilon \rightarrow 0$. 
All other parameters used are the same as in Fig.~\ref{fig1}.}
\label{fig4}
\end{figure}

\begin{figure}
\resizebox{\hsize}{!}{\includegraphics{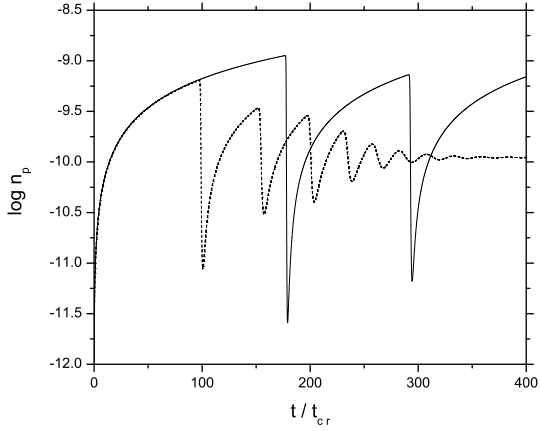}}
\caption{Time evolution of the proton number density for proton injection rate $\log \Qpo = -11.15$ and two
values of the external number density, i.e. $\nex=2$ (solid line) and 
$\nex=4$ (dashed line). The initial conditions for each 
 numerical run are $\nh(0)=\np(0)=0$ and $\ns(0)=\epsilon \rightarrow 0$.  All other parameters used are the same as in Fig.~\ref{fig1}.}
\label{fig5}
\end{figure}

Figure \ref{fig4} shows a two-dimensional plane of the phase space for two different values of the external density. All the other parameters
are kept constant. The corresponding time evolution of the proton distribution is shown in Fig.~\ref{fig5}. 
External photons act as a stabilizing factor for the system, since high enough values lead the system to a steady state.

\section{Enhancing non-linearity with additional processes}
\subsection{Photon-photon absorption by the external source}

In the previous section, in order to avoid the appearance of more terms in the equations and be able to treat
them analytically, we restricted
our analysis to values of $B$ taken from the parameter space shown in Fig.~\ref{Bspace} and
to external photons having energy equal to the threshold energy for proton-photon pion processes.
 If one wishes to use values of the magnetic field more related to astrophysical sources or consider more energetic
external photons, 
 the absorption of hard photons on the external ones
must be taken into account. 

Thus, two more terms appear in the equations of hard and soft photons and the 
corresponding system now is (system S3):
\eqb
\dot{n}_{\rmn p} & = & \Qpo -\frac{\np}{\tp}-\spg \np \nex - \spg \np \ns
\label{eqex1}
\eqe
\eqb
\dot{n}_{\rmn h} & = & -\nh + A \np \nex + A \np \ns - C_{\rmn h} \ns \nh -C'_{\rmn h} \nex \nh
\label{eqex2}
\eqe
\eqb
\dot{n}_{\rmn s} & = & -\ns +C_{\rmn s} \ns \nh + C'_{\rmn s} \nex \nh,
\label{eqex3}
\eqe
where $C'_{\rmn h}=\frac{\sigma_{\gamma \gamma}^{(\rmn {ex})}}{\epsilon_{\rmn o} \eh}$ and
 $C'_{\rmn s}=\frac{\sigma_{\gamma \gamma}^{(\rmn {ex})}}{\epsilon_{\rmn o} \es}$. 
These terms are derived using energy conservation considerations
 as in $\S 3$. The index `ex' is used to remind that the photon-photon 
cross section has a different value depending on whether the absorbing targets 
are the soft or the external photons.

The additional coupling terms make now an analytical study as the one presented in $\S3$ cumbersome. 
Thus, all the results presented in this section
 are derived after solving numerically the stiff set of equations S3. 

\begin{figure}
\resizebox{\hsize}{!}{\includegraphics{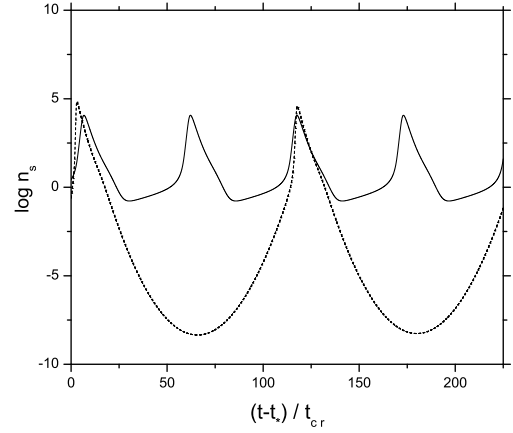}}
\caption{Time evolution of the soft photon distribution $\ns$ in the case where 
hard photons are being absorbed on both the external and the soft 
photons (solid line) and in 
 the case where they are being absorbed only on the latter (dashed line). 
Time is measured with respect to $t_{*}$, that corresponds to the end of a transient phase that 
the system goes through, before it settles
to its periodic state. For reasons of clarity, the transient phase is not shown.
For the parameters used, see text.}
\label{ns-t}
\end{figure}

Our solutions indicate that these extra terms do not change, at least qualitatively,
 the validity of the results of the previous section. Increasing either $\Qpo$ or $\nex$ above a certain value,
 forces the system to fall into a steady state rather than oscillate.
As a first step, we compare two cases: (i) with and (ii) without the extra terms of absorption. 
In both cases we have used  $B=0.7$ G,
$\gp=2 \times 10^7$, $\log \Qpo=-11.15$ and $\nex=2$. 
What differs in the two cases are the external photon energies, that are taken to be (i) 
$\epsilon_{\rmn o} = 100 \gp^{-1} \frac{m_{\pi}}{m_{\rmn e}}$ and (ii) 
$\epsilon_{\rmn o} =\gp^{-1} \frac{m_{\pi}}{m_{\rmn e}}$ respectively.
 Although the system shows is oscillatory in both cases, there is an
important difference that is better displayed when the time evolution
of the soft photon distribution is plotted (see Fig.~\ref{ns-t}). 
At early times, the soft
photon number density evolves in the same way in both cases. However, 
in the first case the number of soft photons in the source does not reach a deep minimum as in the second case.
This can be attributed to the existence of the additional linear 
injection term of soft photons, i.e.,  $+ C'_{\rmn s} \nex \nh$. Since the latter depends only on one time-varying parameter ($\nh$), 
it adds soft photons into the system at a non negligible rate, in contrast to the non-linear term $+C_{\rmn s} \ns \nh$,
that depends quadratically on the time-varying densities.

\begin{figure}
 \resizebox{\hsize}{!}{\includegraphics{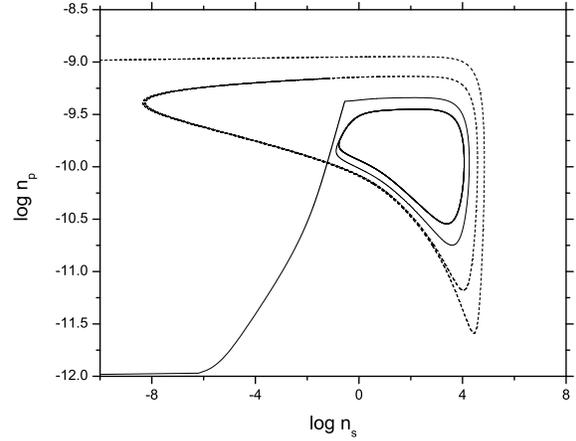}}
\caption{Two dimensional plane $\log \np - \log \ns$ of the phase space in the case where 
hard photons are being absorbed on both the external and the soft 
photons (solid line) and in the case where they are being absorbed only on the latter (dashed line). 
Same parameters used as in Fig.~\ref{ns-t}.}
\label{np-ns}
\end{figure}

This is also reflected on the shape of the
limit cycles in the plane $\np - \ns$ of the phase space, which now appear tighter than before. This is shown in Fig.~\ref{np-ns}.

\begin{figure}
\centering
 \resizebox{\hsize}{!}{\includegraphics{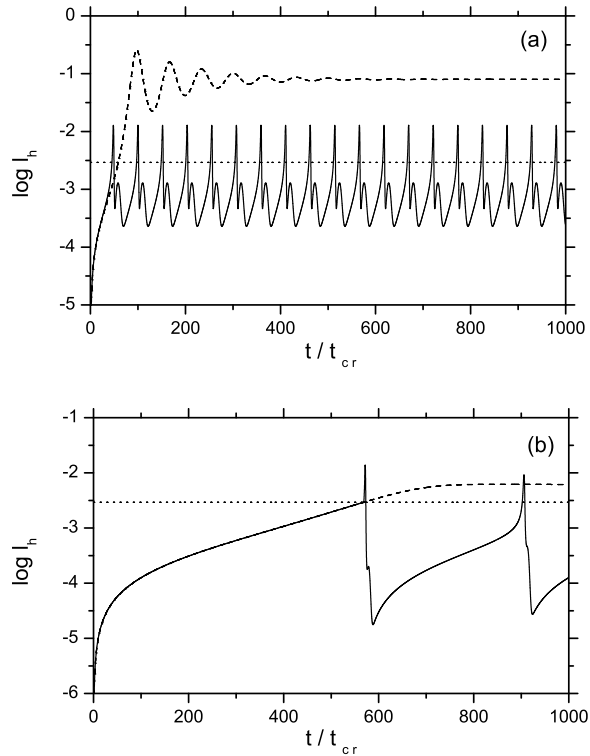}}
\caption{Time evolution of hard photon compactness when (i) $\gamma$-rays are being
absorbed both on external and on 
automatically produced soft photons (solid lines) and 
when (ii)
the non-linear terms of absoprtion are artificially omitted (dashed lines). The dotted line in both panels 
corresponds to $\lhcr$. The cases presented in panels (a) and (b) differ only at the proton injection 
rate, which is taken to be $\Qpo=3.2\times 10^{-11}$  and $\Qpo=4\times10^{-12}$ respectively. The rest of the
parameters used in both panels are the same and equal to: $B=0.75$ G, $\gp=2.65\times 10^7$, $\epsilon_{\rmn o}=10^{-5}$ and $\nex=1$.}
\label{extra}
\end{figure}

As already discussed in section 2.2, the existence of an initial external photon
distribution makes the role of quenching less clear. However, the way the equations of system S3 are written,
 allows us to study separately the linear and non-linear absorption of hard photons by artificially 
deleting the non-linear terms of quenching in the equations of hard and soft photons, i.e.,  $-C_{\rmn h} \ns \nh$ 
and $+ C_{\rmn s} \ns \nh$ respectively. For this purpose we examine two cases, that differ only at the proton injection rate.
Figure ~\ref{extra} shows the hard photon compactness as a function of time for both cases, with panel (a) corresponding to the
case with the larger proton injection rate. 
The horizontal dotted line corresponds to $\lhcr$, the solid line shows $\ell_{\rmn h}$ when both absorption channels
operate, whereas the dashed line shows $\ell_{\rmn h}$ when we ignore the soft photons produced due to non-linear quenching.
 Comparison of the solid and dashed lines leads to the conclusion that the automatic quenching of 
hard photons becomes dominant in the dynamics of the system from the instant that $\ell_{\rmn h} \gtrsim \lhcr$. 
Thus, even if quenching cannot be distinguished from the linear absorption when both operate,
$\lhcr$ still remains an intrinsic property of the system.

Another conclusion drawn from Fig.~\ref{extra} is that even in the absence of automatic quenching
the system can exhibit damped oscillations. These example cases show clearly 
that the large-period limit cycle behaviour 
exhibited when automatic
quenching operates in parallel to the linear one, is replaced
by an expontential growth that saturates, in the case where quenching is artificially omitted.
 On the other hand, the small-period limit cycle behaviour
is replaced by damped oscillations of small amplitude. In other words, the combination of linear and non-linear absorption
of $\gamma$-rays seems to intensify the temporal variability of the system. 
Finally, the absorption of hard photons is more efficient
in the case where both the linear and non-linear channels of absorption operate than in the case where
hard photons are being absorbed only on the external photon population. This can be deduced from
 the fact that, in the former case the averaged $\ell_{\rmn h}$  over a period 
is suppressed by at least one order of magnitude (see solid and dashed lines in panel (a) of Fig.~\ref{extra}).

\begin{figure}
\centering
 \resizebox{\hsize}{!}{\includegraphics{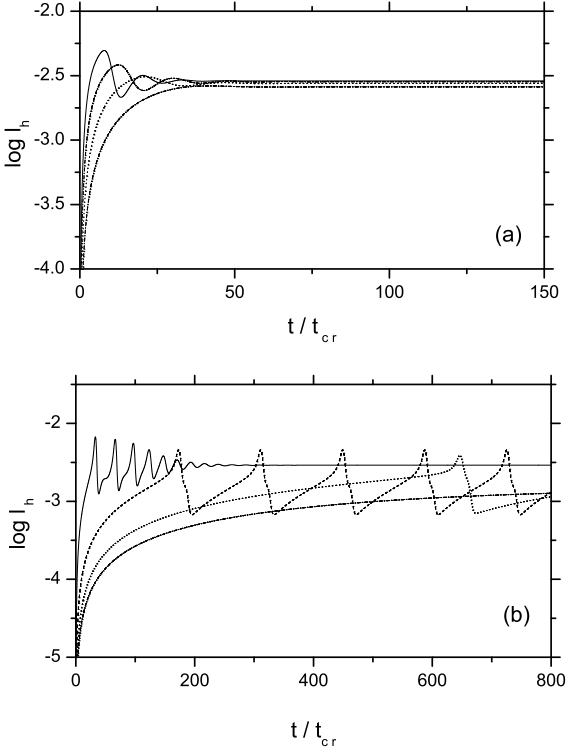}}
\caption{Hard photon compactness as a function of time for two cases with $\tau_{\rmn {ex}}=1.3$ (panel a) and $0.13$ (panel b).
In both panels the different light curves are obtained by increasing the proton injection rate.
The transition from the linear to the non-linear operating
regime of the system is clearly seen.
Specifically, for panel (a) we have used: $\Qpo=10^{-11}$ (dash-dotted line), $2\times 10^{-11}$ (dotted line), 
$4 \times 10^{-11}$ (dashed line) and $8 \times 10^{-11}$ (solid line). The corresponding values for panel (b) are:
$\Qpo=10^{-12}$ (dash-dotted line), $1.6\times 10^{-12}$ (dotted line), $4 \times 10^{-12}$ (dashed line) and  
$2.5\times10^{-11}$ (solid lines). The rest of the parameters used are the same for both panels:
$B=0.75$ G, $\gp=2.65\times 10^7$ and $\epsilon_{\rmn o}=10^{-3}$. 
}
\label{tau-ex}
\end{figure}

Two operating regimes of the system, each of them having its own properties, 
can be defined, depending on the relative strength of the two absorbing channels:
 (i) a linear and (ii) a non-linear one.
While being in the linear regime, $\ell_{\rmn h}$  grows at first exponentially and eventually saturates. 
The transition from the linear to the non-linear operating regime depends on the external number density or equivalently on
the optical depth $\tau_{\rmn {ex}}$ for absorption of hard photons on the external ones. In our analysis,
 $\tau_{\rmn{ex}}$ is simply given by
\eqb
\tau_{\rmn {ex}} = \sgg(\eh \epsilon_{\rmn o}) \nex(\epsilon_{\rmn o}).
\eqe
For small optical depths, i.e., $\tau_{\rmn {ex}} \ll 1$, this transition is abrupt
in the sense that the system changes its temporal behaviour completely. From the instant 
that automatic quenching becomes  the dominant channel for absorption, the system
  exhibits limit cycles of large period and amplitude.
On the other hand, for large optical depths, i.e., $\tau_{\rmn {ex}} \gtrsim 1$, the transition is smooth, 
since the system shows no limit-cycle behaviour.
The non-linearity in the temporal behaviour becomes evident by damped oscillations that reach a steady state in a few crossing times.
Thus, the flaring behaviour of the system can be severely suppressed whenever the optical depth for absorbing $\gamma$-rays on external photons
is large. Panels (a) and (b) of Fig.~\ref{tau-ex} show $\ell_{\rmn h}$ as a function of time for two cases,
 where $\tau_{\rmn {ex}}=1.3$ and $0.13$ respectively. In each panel, lines of different type denote different proton injection rates. 
The lightcurves depicted with dashed-dotted lines in both panels are obtained while the system operates in its linear regime.
The dotted lightcurves exemplify the transition to non-linearity, which is more abrupt for the case in panel (a) than 
the corresponding in panel (b). 

\begin{figure}
 \resizebox{\hsize}{!}{\includegraphics{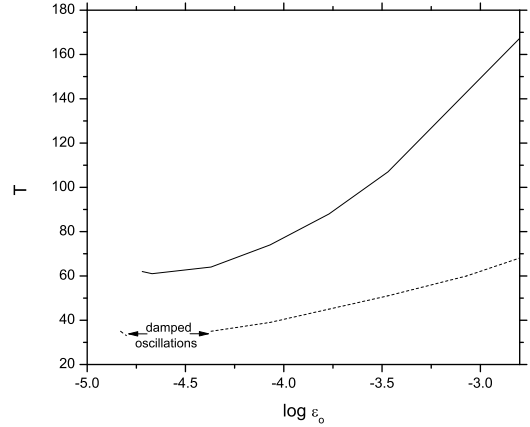}}
\caption{Dependence of the  period T on the energy of the external
 photons $\epsilon_{\rmn o}$ for number densities $\nex=1$ (solid line) and $\nex=3$
(dotted line). 
The rest of the parameters used are: $\Qpo=10^{-10}$, $B=3.57$ G, $\gp=9\times 10^6$ and $\epsilon_{\rmn o}=8\times10^{-5}$.}
\label{T-ex}
\end{figure}

A new feature that appears through the study of system S3 is 
the dependence of the period T, if this exists, on the energy of the
external photons $\epsilon_{\rmn o}$. This is exemplified in Fig.~\ref{T-ex}, where 
the period of the oscillations varies with $\epsilon_{\rmn o}$
 almost like $T \propto 1/ \sgg(\epsilon_{\rmn o})$ -- see eq.~(\ref{sgg}). 
The minimum period is found at $\epsilon_{\rmn o}$ that corresponds to the maximum value of the cross section
for photon-photon absorption; for this value the absorption of hard photons becomes
most effective. 
The dashed line shows the dependence of the period on $\epsilon_{\rmn o}$ for a higher density of external photons.  
In this case a gap appears for values of $\epsilon_{\rmn o}$ that correspond to high values of the cross section around its peak. 
The evolution of the system there, is characterized
by damped oscillations that lead eventually to a steady state. 
 This result is to be expected within our analysis of $\S 3$. We remind that the system passes through
well-defined stages, as one of the parameters $\Qpo$ or $\nex$ increases:
 oscillations with large period $\rightarrow$ oscillations with small period $\rightarrow$
 damped oscillations leading to a steady state.

The fact that we find a clear analogy between the period and the inverse of the cross section for photon-photon absorption
is a direct consequence of the simplifications we have made in the problem so far. However we note that, if were to relax our assumptions, 
i.e., use full expressions for the cross sections and emissivities, and treat
the problem numerically, we would still have retained the basic conclusions of Fig.~\ref{T-ex}.

We have also found that the period of the limit cycles depends not only on $\epsilon_{\rmn o}$ but also on other
parameters, that affect the value of the cross section of photon-photon absorption even indirectly, as the magnetic field strength or/and the 
proton energy
(see eqs. (\ref{gepion}), (\ref{eh}) and (\ref{es})).  The trend is the same as the one shown in Fig.~\ref{T-ex}, 
where $\epsilon_{\rmn o}$ in the horizontal axis should be replaced by the corresponding varying parameter.

\subsection{Inverse Compton scattering}

Thus far, we have assumed that the created pairs from photon-photon absorption act as an agent,
 transfering the energy from hard 
to soft photons through synchrotron radiation.
 However, if the compactness of soft photons becomes comparable or 
 larger than the compactness of the magnetic field, i.e., $\ls \gtrsim \lB$, 
then there are two cooling channels
for the secondary electrons: (i) the `synchrotron' one that
results to the production of soft photons $\es$ and (ii) the
`inverse Compton' (ICS) one that results to the production
of high energy photons $\epsilon_{\rmn {ics}}$ -- note that
in general $\epsilon_{\rmn {ics}} \neq \eh$.
Thus, the energy lost by hard photons
 is only partially injected to the soft photon population. Because of this, the 
constant $C_{\rmn s}$ of the injection term in eq.~(\ref{eqex3}) should be replaced by
\eqb
C_{\rmn s}^{\rmn {eff}} = C_{\rmn s}\frac{\lB}{\lB+ 3 \ls (1+4\es \gamma_{\rmn e})^{-3/2}},
\label{ICSeff}
\eqe
where $\gamma_{\rmn e} = \eh/2$ and the multiplication factor of $\ls$ accounts approximately for
the Klein-Nishina cutoff effect up to $\es \gamma_{\rmn e}\lesssim 10^4$ \citep{moderski05}.
Inspection of system S3 together with the expression (\ref{ICSeff}) shows that the 
inclusion of ICS adds more non-linear terms to the problem.

It is beyond the scope of the present work to proceed to a semi-analytical study of the above system.
However, it is worth mentioning some qualitative effects of ICS on the dynamics of the system. 
In general, ICS acts as a damping term for soft photons whenever
$\ls \gtrsim \lB$.

\begin{figure}
\centering
 \resizebox{\hsize}{!}{\includegraphics{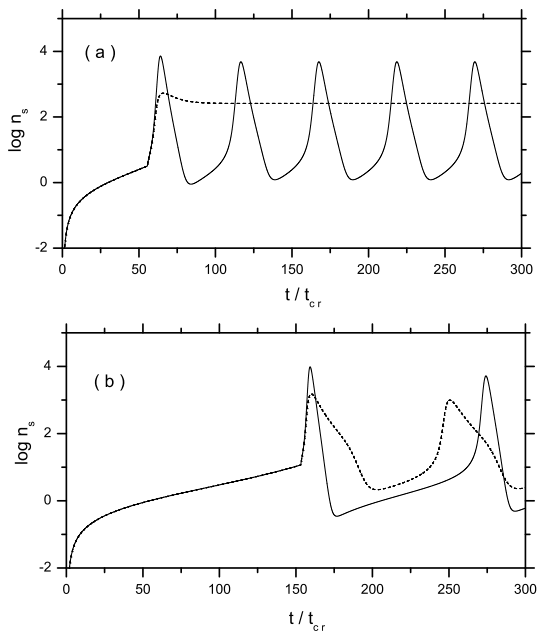}}
\caption{Soft photon density $\ns$ as a function of time for proton Lorentz factors 
$\gp=2.1\times10^7$ and $2.65\times10^7$ (panels a and b respectively), derived for two cases:
 (i) ICS is artificially switched-off (solid
lines) and (ii) ICS is taken approximately into account (dashed lines). Other parameters used for this plots are:
$B=0.75$ G, $\log \Qpo=-11.15$, $\nex=2$ and $\epsilon_{\rmn o}=\gp^{-1} \frac{m_{\pi}}{m_{\rmn e}}$.}
\label{ics}
\end{figure}

Let us assume first, that we artificially
switch-off ICS, and find a set of parameters that lead our system to a limit cycle
behaviour as discussed in the previous section. If we keep the parameters fixed to these values and switch-on ICS, 
then there are three possible ways for the evolution of the system : 
\begin{enumerate}
 \item [(i)]
The limit cycle behaviour remains, although the system oscillates
with a smaller period.
\item[(ii)]
The limit cycle behaviour is
damped and the system finds its steady state after a number of oscillations.
\item[(iii)]
The system falls quickly into a steady state before showing any oscillations.
\end{enumerate}
The resulting behaviour of the
system depends on the ratio $\ls / \lB$ and on whether or not the scatterings occur in the Klein-Nishina regime.
 Figure \ref{ics} exemplifies the above remarks. 
The solutions shown in Fig.\ref{ics} are obtained after integrating the system of equations
S3 and incorporating ICS in the approximate way described in this section.
In both cases, the system starts with $\ls \ll \lB$
but eventually reaches a state where $\ls \gtrsim \lB $. 
The difference between the cases above
 is the parameter $x_{\rmn {ics}}=\es \gamma_{\rmn e}$, that denotes
how deep into the Klein-Nishina regime the scatterings occur.
Cases presented in panels (a) and (b) correspond to values $x_{\rmn {ics}} = 7.6$ and $87$ respectively.
 In the former case, the damping effect of ICS is evident, whereas in the latter case 
 the system's evolution is not much affected because of the suppression of the scatterings. A small decrease in the period and in the
amplitude of the oscillations is however evident.
 In the following section, where we treat numerically the full problem, we present 
a case that exemplifies the effects of ICS.

\section{Numerical approach}
All of our previous results were verified in an independent way by using the numerical code described in
MK95 after selectively omitting various processes, as to make the code analogous to the systems
described in the previous section.

We proceed next to solve numerically the full system of eqs. (\ref{generic}).
Our aim is to present only some characteristic examples which will support our previous analysis and
show also the effects of the processes we have neglected in our analytical approach, 
most notably of inverse Compton
scattering, on the dynamical behaviour of the system.
We leave a comprehensive search of the parameter space for a future paper, where the interplay
between proton injection and external photons will be fully addressed.

We have used the numerical code described in MK95 and MPK05 which has been
updated as to make use of the full rates for the 
secondary electron-positron pairs and photon production in photopion
interactions as obtained from the SOPHIA Monte Carlo code \citep{mueckeetal00}.
Details on these will appear elsewhere -- 
Dimitrakoudis et al. in preparation; see also \cite{dimitrakoudisetal12}.
Therefore the updated version of the code can treat accurately the two major hadronic
processes, i.e., photopair and photopion, in addition to the leptonic ones. 
Given the difficulty 
that these two processes pose in modelling, we consider this as a major improvement.
 
We solve therefore three coupled equations, for protons, electrons and photons including all
relevant processes between the three species -- note that in the numerical code there is no need
to treat the hard and soft photons through separate equations. Another difference with
the analytical treatment is that in place of the external photons we use the photons produced 
by the proton synchrotron radiation. This was done because this process
can produce the seed photons self-consistently without the need of introducing more free parameters. Furthermore,
the proton synchrotron radiation, for magnetic fields of $\sim 10$ G and protons with Lorentz factors $\gp \gtrsim 10^6$ is produced
mainly in the soft energy range of the spectrum and cannot/should not be neglected.

\begin{figure}
\centering
 \includegraphics[width=9.5cm, height=9.cm]{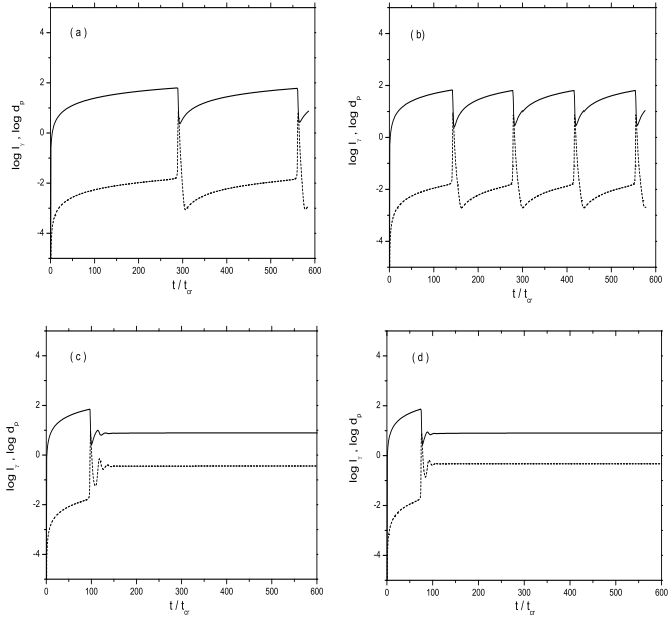}
\caption{Time evolution of the system for four different proton injection rates or equivalently proton compactnesses,
starting with $\ell_{\rmn p}^{\rmn {inj}}=4.7\times10^{-5}$ in panel (a). 
In panels (b) to (d) $\ell_{\rmn p}^{\rmn {inj}}$ is increased over its previous value by an integer multiple of its initial
value. The rest of the parameters used for this plot are: $B=10$~G, $R=10^{16}$ cm and $\gp=3\times 10^7$.
Solid and dashed lines show $d_{\rmn p}$ and $\ell_{\gamma}$ respectively, 
where $d_{\rmn p} = \int \diff \gamma_{\rmn p} \gamma_{\rmn p} \np (\gamma_{\rmn p})$.
}
\label{num1}
\end{figure}

Fig.~\ref{num1} shows four cases that differ only in the proton injection rate. Thus for panels (b),
(c) and (d)  $\Qpo$ was increased by a factor of 2, 3 and 4 respectively over its corresponding
value of panel (a).
The latter one was chosen in such a way as to make the system exhibit large period limit cycles.
The period starts decreasing with increasing $\Qpo$
-- as a matter of fact the period gets exactly to a half of its previous value as $\Qpo$ is 
increased by a factor of 2. This behaviour degenerates into a damped oscillation --
steady state mode with increasing $\Qpo$ (panels c and d). This is exactly the behaviour we found in
our analytical treatment -- see Fig.~\ref{figclose}. Therefore, despite the plethora of the physical processes 
introduced, the feedback loop still operates.

\begin{figure}
\centering
 \includegraphics[width=8.cm, height=7.cm]{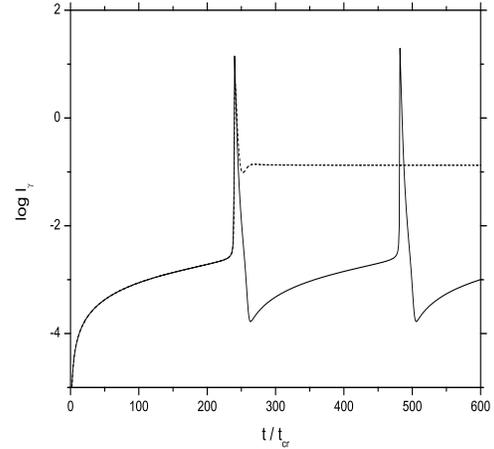}
\caption{Photon compactness $\ell_{\gamma}$ as a function of time for 
two cases where, (i) inverse Compton scattering is taken into
account (solid line) and (ii) it is artificially switched-off. 
 The parameters used for this plot are: $B=3.2$ G, $R=10^{16}$ cm and $\gp=3\times 10^7$, 
$\ell_{\rmn p}^{\rmn {inj}}= 7 \times 10^{-5}$.
}
\label{num2}
\end{figure}

For low values of the magnetic field, ICS acts as a friction
mechanism stabilizing the system and letting it reach quickly steady state. Fig.~\ref{num2}
shows an example where we ran the code for two cases. In the first, all processes
were taken into account (dashed line), whereas in the second case, ICS was artificially switched-off (solid line).
In the former case the system, after an initial peak, falls quickly into a steady state.
In the latter it shows the limit cycle behaviour.  The two cases are identical
until the time when the compactness of soft photons becomes large and cannot be neglected any further in the electron cooling. 
 This occurs around the time that the first peak in $\ell_{\gamma}$ appears, 
where the soft photon compactness becomes by a factor of $\sim 6\times 10^3$
higher than the magnetic one.
It is worth comparing the result shown in Fig.~\ref{num2}
 with the one plotted in panel (a) of Fig.~\ref{ics}, where ICS was taken into account in an approximate
way. In both cases the qualitative results are the same. 
 If we use a higher value of the magnetic
field for the example case shown in Fig.~\ref{num2}, we find that ICS alters the periodic behaviour only by decreasing the period. 
The temporal behaviour we find in this case,
 can be very well described by the corresponding one shown in panel (b) of Fig.~\ref{ics}. 
Suppresion of ICS because of Klein-Nishina cutoff effects does not play an important role in this case, since
we do not assume monoenergetic electron and photon distributions. Thus, most of the scatterings occur in the Thomson regime
and Klein-Nishina cutoff effects are now small corrections. 
Finally, we note that the limit cycle behaviour of the system shown in Fig.~\ref{num1} remains intact, although
ICS was taken into account, since for the parameters used, 
the magnetic energy density is always larger than the
soft photon one. 

\section{Relevance to astrophysical sources}
We turn next to examine the ideas presented in the previous sections
 in the context of possible applications of astrophysical interest.
Whenever the system operates in the subcritical regime or the $\gamma$-rays
are being absorbed mainly through the linear absorbing channel (see \S 4.1),
our steady state spectra are similar to those presented in the literature -- 
for example see \cite{boettcher07}. 
On the other hand, if the system becomes supercritical and the absoprtion of $\gamma$-rays is strongly
non-linear,
then new possibilities open for astrophysical applications. These, 
according to \S 3 and \S4, can be summarized in the following:

\begin{enumerate} 
\item 
All the components of the hadronic system exhibit an intrinsic variability with a well defined
period, although
the source is stationary. \\
\item
The system reaches a steady state after going through a damping oscillatory phase. In this
case a soft photon component emerges since a significant fraction of the energy stored in
protons is transferred to lower energy photons via quenching of the $\gamma$-rays.
 At the same time the hard photon compactness reaches a limiting maximum value.
\end{enumerate}

Both cases can, in principle, have relevance to astrophysical sources emitting in high energies, such as AGN.
Time variability is a defining property of blazars and, in most cases, 
it shows a quite complex pattern (e.g., \citealt{mukherjee99, aharonian07}). 
We note that, even if the observations seem to
contradict our results (see point (i) above), we have found that even small
amplitude variations of the proton injection rate can lead to much more
complicated time profiles than the ones presented so far. This is a promising topic.
Therefore, a detailed study of the system towards this direction is required
 and is going to be the subject of a future work.

In the rest of this section we will focus on the second point. 
We will show specifically how we can apply our results 
in order to set an upper limit to parameter values 
 used in the modelling of AGN multiwavelength spectra. This can be seen as
an extension of the application presented in PM11.

The luminous blazar 3C 279 is a good example.
A recent comprehensive review of observations can be found in \cite*{boettcher09}.
Here we will mainly focus on the 2006 campaign, which
discovered the source at VHE $\gamma$-rays, showing a high TeV flux (Albert et al. 2008), while 
the X-rays were at a much lower level.

Let us consider a spherical source of radius R moving with a Doppler factor
$\delta$ with respect to us and containing a magnetic field of strength B. 
We further assume that ultra-relativistic protons with a power law distribution of index $s$
are being constantly injected into the source with a rate given by
\eqb
\tilde{Q}_{\rmn p}=\tilde{Q}_{\rmn {po}} \gp^{-\rmn s} H(\gp-\gammamin) H (\gammamax-\gp),
\eqe
where $\gammamin$ and $\gammamax$ are the lower and upper limits of the injected distribution respectively. 
 $\tilde{Q}_{\rmn {po}}$ is  the normalization constant and is also directly related 
to the proton injection compactness as:
\eqb
\ell_{\rmn p}^{\rmn {inj}} & = & \tilde{Q}_{\rmn {po}} m_{\rmn p} c^2 \frac{\sth R }{3 \tcr} \frac{\gammamin^{-\rmn s+2}-\gammamax^{\rmn s+2}}{2-s}
\eqe
or in terms of dimensionless quantities
\eqb
\ell_{\rmn p}^{\rmn {inj}} = \frac{1}{3} \Qpo   \frac{\gammamin^{-\rmn s+2}-\gammamax^{\rmn s+2}}{2-s}. 
\eqe
The transition of the system from the sub - to the super-critical regime
 can be better seen if initially there are  no soft photons present in the source. 
For this, we try to fit only the TeV emission by considering a narrow power law 
proton energy distribution and no primary electron population.  Thus, the multiwavelength
spectrum is purely the result of proton primary and secondary emission. 

\begin{figure}
\centering
 \resizebox{\hsize}{!}{\includegraphics{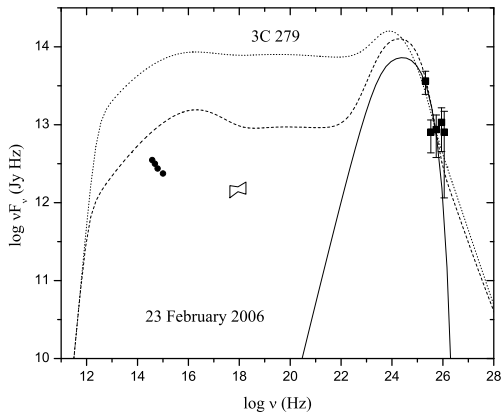}}
\caption{Multiwavelength spectra of 3C 279 obtained in the context of a pure
hadronic model for $R=3\times 10^{16}$ cm, $B=40$ G, $\delta=20$, $\gammamin=5 \times 10^8$, $\gammamax=5\times 10^9$,
$s=2.2$ and three proton injection compactnesses: $\ell_{\rmn p}^{\rmn {inj}}=10^{-5}$ (solid line),
$2\times 10^{-5}$ (dashed line) and $4\times 10^{-5}$ (dotted line). 
The symbols represent the observational data from February 2006.}
\label{fit}
\end{figure}

Figure \ref{fit} shows the multiwavelength spectra obtained using the numerical code described in
\S5 for $R=3\times 10^{16}$ cm, $B=40$ G, $\delta=20$, $\gammamin=5 \times 10^8$, $\gammamax=5\times 10^9$, $s=2.2$ and for three
values of the proton injection compactness starting  with $\ell_{\rmn p}^{\rmn {inj}}=10^{-5}$ (solid line) and  increasing
it over each previous value by a factor of two.
For this set of parameters the produced $\gamma$-rays
lie in the GeV-TeV regime. The spectrum for the lowest value
of $\ell_{\rmn p}^{\rmn {inj}}$ (solid line) is obtained while the system is subcritical,
 and it is the only one that can give an
acceptable fit of the TeV emission while at the same time does not violate the optical and X-ray observations. 
We see that an increase of $\ell_{\rmn p}^{\rmn {inj}}$ by just a factor of two drives the system into the 
supercritical regime. The onset of supercriticality is acompanied by the emergence of a 
soft emission component (dashed line), that becomes dominant for an even higher $\ell_{\rmn p}^{\rmn {inj}}$ (dotted line).
The over-production of soft photons even in the second case is excluded directly by the observations, setting 
an upper limit to the proton injection compactness  (for the specific example, $\ell_{\rmn {p,max}}^{\rmn {inj}}=10^{-5}$).

If we were to use a lower value of $\delta$ to obtain the fit in the TeV energy range
while keeping fixed the magnetic field strength, 
we would require a higher value of $\ell_{\rmn p}^{\rmn {inj}}$, since 
$L_{\rmn {obs}} \propto \delta^4 L_{\rmn {int}}$, where
$L_{\rmn {obs}}$ and $L_{\rmn {int}}$ are the luminosities in the observer's and in the comoving frame respectively.
This choise of parameters would drive the system deep into the supercritical regime violating the optical and
X-ray observations.
 PM11 have used similar arguments to set constraints on the Doppler factor $\delta$ by
using an \textit{ad-hoc} $\gamma$-ray injected luminosity. Here we go one step further since the injected $\gamma$-rays
are related to a physical production mechanism. 
Thus, in this case we can set a limit on both $\delta$ and $\ell_{\rmn p}^{\rmn {inj}}$.
 Moreover a potential flaring event observed only in the GeV part of the spectrum
 could not be fitted by just increasing 
the proton injection luminosity, since this increase would affect also
the optical and X-ray part of the spectrum, as the example of Fig.~\ref{fit} suggests.

Thus, the effects of the underlying feedback mechanism 
can prove to be useful in setting lower limits to parameters, such as the 
Doppler factor $\delta$.
A systematic search of the parameter space is however out of the scope of the present paper.

\section{Discussion}
Hadronic models have been extensively used for explaining AGN non-thermal
emission while recently thay have been applied, as well, to other compact objects.
An interesting, but overlooked property of hadronic systems is their dynamical 
behaviour, which results from some underlying feedback mechanisms. In the present paper
we have isolated and studied analytically one such loop that involves proton-photon pion
production and photon quenching (SK07; PM11) of the produced $\gamma$-rays.
Gamma-ray quenching results in automatic production of soft photons which
then feed back producing more proton-photon cooling via pion production.

We have remarked that if protons are considered stationary in the source
-- an often made assumption, 
there are parameter space regimes which are characterized by an exponential 
growth of internally produced photons, making the system inherently unstable
since the condition for proton stationarity is violated by the
losses caused by the runaway photons -- here the analogies 
to the `Compton catastrophe' of leptonic plasmas are evident. 
The kinetic equation approach,
which allows proton cooling to be explicitly taken into account,
 is suitable for studying in this case, the properties of the system.
Protons, secondary electrons and 
photons, i.e., the main three components of the system,
 can be described by three coupled partial integro-differential equations. This
`kinetic equation treatment'  has
many advantages, as it is both energy conserving and
time dependent.

In order to simplify the system of equations and make an analytical
treatment possible we made a number of assumptions.
As a first step 
we have retained only simplified expressions of
the key processes by using $\delta$-functions for the different particle
distributions appearing in the problem. Furthermore, we used approximate expressions
for the cross sections -- see eqs.~(\ref{spg0}) and (\ref{sgg}). 
However, one of the major simplifying assumptions
we made was the elimination of the electron kinetic equation from the system of eqs.~(\ref{generic}). 
The rationale
for this is that the electron cooling timescale, for typical values of the magnetic field and electron Lorentz factors,
is much smaller than the crossing time of the source. Thus, electron cooling is considered to be instantaneous.

Ignoring absorption of the hard photons
on the external ones, we have found that
an increase 
of the proton injection rate leads to an analogous increase of the proton density 
and of the hard photon luminosity  
resulting from photopion. If the latter does not
get to a high value as to trigger quenching, the system 
is linear and reaches a steady state. If, however, the combination
of the initial parameters is such that leads
the hard photons past the quenching
threshold, then the latter are automatically absorbed 
and the soft photons which are spontaneously produced serve
as targets for extra proton cooling. 
In this case we have shown analytically by performing
an eigenvector/eigenvalue analysis,
that for hard photon densities close but above critical
the system goes through a limit cycle behaviour
of the prey-predator type. For even higher hard photon densities,  
the system reaches  a steady state which is achieved after
protons and photons exhibit a series of
damped oscillations. This steady state occurs at very different values
from the ones achieved
while the system is subcritical and this discontinuity is another 
indication of the system's supercriticality.
It is interesting to note that duty-cycle behaviour in hadronic systems
was reported by \cite{stern91} using a Monte Carlo code and by MPK05 using
a kinetic equation approach. However both papers were numerical
and the authors, while giving ample physical reasoning for the
behaviour, fell short of presenting a mathematical proof. Here for
the first time we present such an interpretation and show beyond doubt, that
hadronic systems can indeed exhibit the aforementioned behaviour.

As a next step we have shown semi-analytically 
that when absorption of the produced hard photons
on the external ones is included, then the behaviour of the system depends also
on the corresponding optical depth. If this takes low values,
then the system behaves as described above since the non-linear
processes continue to play a dominant role in the dynamics of
the system. If, on the other
hand, the optical depth takes high values, the systematic 
depletion of hard photons tends to stabilize
the system which reaches a steady state after it goes through a damped
oscillation mode, i.e., no limit cycle behaviour was found in this case.
It is interesting however to note that, even in this case, 
the hard photons cannot reach a steady state above the
critical value of quenching. This can only mean that quenching remains 
a fundamental intrinsic property of the system.

Furthermore, if we are to add 
more physical  processes to the system, these  
tend to stabilize it as they redistribute
part of the radiated energy away from the operating feedback loops.
For example, inverse Compton scattering can act as a competing 
energy loss mechanism for electrons to synchrotron radiation. If it dominates,
then the system changes behaviour moving faster to a steady state.
However for strong enough B-fields, we found that the system retains the analytically derived 
 properties.

The above results, were also verified by 
using a numerical code where
the full expressions
for the emissivities of the various radiative processes
 and for the various relevant
cross sections were used. 
While the details change from our simplified analytical approach, 
we were able to 
confirm qualitatively our analytical results which predict the transition from the subcritical
linear regime to the supercritical oscillatory one 
with increasing proton density. 
In addition, we were able to verify
the role of other processes, like inverse Compton scattering, which we had taken in our analytical treatment
only approximately into account. 
We note that the qualitative analogies between the results of the two treatments
justify also \textit{a posteriori} and in an independent way
the validity of our simplifying assumptions.

 Our results indicate that higher proton injection rates tend to push
the system into the non-linear regime with the external photons acting more as catalysts; 
on the other hand, higher external photon densities act on the opposite direction and tend
to linearize the system. Preliminary numerical calculations (Dimitrakoudis et al. -- in preparation)
show that this trend remains, if one is to replace the $\delta$-function distributions used
in the present treatment with the more astrophysically relevant power-laws. 

Finally, as an example of astrophysical interest, we have used the numerical code described in
\S5, which can treat self-consistently both the non-linear development of EM cascades and proton cooling, in
order to make a fit to the TeV emission of blazar 3C~279. We have shown
that acceptable fits can be obtained only for high values of the Doppler
factor ($\delta \gtrsim 20$), given that
typical values for the magnetic field 
 in the context of hadronic models 
are considered (B $\simeq$ 10 - 60 G).
However, this is only an indicative example of the potential
applications of the model. Obviously one needs to thoroughly study the parameter space
before providing exact fitting values. Another potential direction is the study of the
inherent variability signatures of the system in cases where the source itself is variable.

The supercriticality related to the feedback mechanism studied in the present paper
is by no means the only one that can occur in hadronic systems. 
If one was to replace the photopion interactions by another 
production mechanism of $\gamma$-rays, e.g., proton synchrotron radiation,
and study the same feedback loop outlined in \S1, one would again find 
that the system enters a supercritical regime. However, the
parameter values that would enable the transition from the subcritical to 
the supercritical regime,  as well as the
the details concerning the transition itself, would differ from those shown in the present work,
due to diferrent cross sections, emissivities and energy threschold criteria.
Furthermore, we cannot exclude that even other loops
operate as well in a hadronic system -- see \cite{kirkmast92}, \cite{dimitrakoudisetal12}. 
Actually, more than one of the different feedback mechanisms can operate simultaneously in a `real' hadronic
system. Because of this, a comprehensive search of its parameter space  
demands the use of the numerical code described in \S5 and
it is going to be the subject of a future work.
At any rate, we have shown that hadronic models constitute one more example
in the growing list of dynamical systems and as such they need to be 
further investigated.

.

\section*{Acknowledgments}

We would like to thank Drs. R. J. Protheroe and A. Reimer for
making available the SOPHIA results to us and S. Dimitrakoudis for incorporating
them in the numerical code. We would like to thank also Drs. N. Vlahakis and C. Efthymiopoulos 
for useful comments and discussions.
This research has been co-financed by the European Union
(European Social Fund - ESF) and Greek national funds through
the Operational Program `Education and Lifelong Learning' of 
the National Strategic Reference Framework (NSRF) - Research
Funding Program: Heracleitus II. Investing in knowledge society
through the European Social Fund.

\appendix

\section[]{Stability analysis of the trivial stationary solution}

Consider the system (S)
\eqb
\dot{x} & = & -x + A  \nex z + A z y - C_{\rmn h} y x \\
\dot{y} & = & -y +C_{\rmn s} y x\\
\dot{z} & = & \Qpo -\frac{z}{\tp}-\spg  \nex z - \spg z y
\eqe
where variables $(x,y,z)$ stand for $(\nh,\ns,\np)$. The fixed points
 of the system, $P_i(x_0^{(i)},y_0^{(i)},z_0^{(i)})$ with $i=1,2,3$
can be found by setting $\dot{x}=\dot{y}=\dot{z}=0$. The first fixed point is nothing more than the trivial steady state
solution of the system in the case of no quenching:
\eqb
x_0^{(1)}& = & A \nex \frac{\Qpo}{G_{\rmn p}} \\
y_0^{(1)} & = & 0 \\
z_0^{(1)} & = & \frac{\Qpo}{G_{\rmn p}}.
\eqe
The remaining fixed points have $x_0^{(2),(3)}=1/C_{\rmn s}$ and
 $z_0^{(2),(3)}=\frac{\Qpo}{G_{\rmn p}+\spg y_0^{(2),(3)}}$, 
where $y_0^{(2),(3)}$ are the
real roots of the equation
\eqb
y^2+y\! \left(\! \frac{G_{\rmn p}}{\spg}+\! \frac{1}{C_{\rmn h}}-
\frac{\Qpo A}{\spg} \frac{C_{\rmn s}}{C_{\rmn h}}\right)\! 
+\!\!\frac{G_{\rmn p}-\Qpo A \nex C_{\rmn s}}{\spg C_{\rmn h}}\!=\!0.
\eqe
We make the convention that $y_0^{(2)}$ is the positive real root that has physical meaning.
It is interesting to examine the behaviour of the system when it is slightly perturbed by its steady state, 
i.e., $x=x_0^{(1)}+x', y=y', z=z_0^{(1)}+z'$. For this we linearize system (S) with respect to the perturbed quantities:
\eqb
\dot{x'} & = & -x'+(A z_0^{(1)} -C_{\rmn h} x_0^{(1)}) y' + A \nex z' \\
\dot{y'} & = & (-1+C_{\rmn s} x_0^{(1)}) y' \\
\dot{z'} & = & -\spg z_0^{(1)} y' -G_{\rmn p} z'.
\eqe
For the second equation that is not coupled to the other two we find an exponential growth or decay $y'(\tau)=y'(0) e^{\rmn s \tau}$
depending on the sign of  $s=-1+C_{\rmn s} x_0^{(1)}$. First suppose that $s<0$. 
$y' \rightarrow 0$ holds for $\tau \gtrsim 1/s$ and the three linearized
equations degenerate to two. The matrix of the two dimensional system is
\begin{displaymath}
\textrm{\sffamily{M}}_1 = \left( \begin{array}{c c}
                       -1 & A \nex \\
		        0 & -G_{\rmn p}
                      \end{array} \right) 
\end{displaymath}
with determinant $\Delta(\textrm{\sffamily{M}}_1) = G_{\rmn p} >0$ and 
trace $\textrm{Tr(\sffamily{M}}_1) = -1-G_{\rmn p} <0$. Thus, in this case 
the point $P_1$ is stable. 
This simple analysis does not apply in the case of $s>0$. However, it can be easily shown that both
 $z'$ and $x'$ are $\propto e^{\rmn s\tau}$ for large enough times. Thus, in this case all the perturbed quantities grow with time
and $P_1$ can be characterized as unstable.

\section[]{Stability analysis of the non-trivial stationary solution}
In general, the topology of a vector field near its fixed points can be studied through 
 the Jacobian matrix of the vector field at the corresponding points. Specifically, the classification of
 fixed points in different types is made by an eigenvalue/eigenvector analysis of the Jacobian matrix.
This analysis is widely used for two-dimensional vector fields, leading to a few types of fixed points. This is not
the case for three dimensional systems, where the classification of the fixed points in types is more complicated.
In our work we have adopted the classification presented in \cite{theisel03}.

Here we apply the eigenvalue analysis to the three dimensional vector field $\mathbf{v}= (\dot{n}_{\rmn h},\dot{n}_{\rmn s},\dot{n}_{\rmn p})^T$.
We consider the second non-trivial stationary solution of the system S2. Thus,
the set of equations (S2) after linearization at the point $P_2$ can be written in the form
\begin{displaymath}
\begin{array}{ccc}
\left( \begin{array}{c}
        \dot{n}_{\rmn h} \\
	\dot{n}_{\rmn s}\\
	\dot{n}_{\rmn p}
       \end{array}\right) &\!\!\! = \! \textrm{\sffamily{M}}_2 & \!\!\!\!\left( \begin{array}{c}
        \nh \\
	\ns\\
	\np
       \end{array}\right) \\
\end{array}
\end{displaymath}
where the matrix \sffamily{M}$_2$ is given by 
\begin{displaymath}
\textrm{\sffamily{M}}_2 = \left( \begin{array}{c c c}
                       -1-C_{\rmn h} y_0^{(2)} & A z_0^{(2)} - C_{\rmn h} x_0^{(2)} & A(\nex+y_0^{(2)}) \\
		        C_{\rmn s} y_0^{(2)} & 0 & 0 \\
			0 & -\spg z_0^{(2)} & -(G_{\rmn p} + \spg y_0^{(2)})
                      \end{array} \right).
\end{displaymath}
The eigenvalues are the roots of its characteristic polynomial 
\eqb
P(\lambda)=\lambda^3+a_1 \lambda^2 +a_2 \lambda +a_3,
\label{plamda}
\eqe
where
\eqb
a_1 & = &-\textrm{Tr(\sffamily{M}}_2)\\
a_2 & = & \left(1+C_{\rmn h} y_0^{(2)}\right)\left(G_{\rmn p}+\spg y_0^{(2)}\right)- \nonumber \\
    &-& C_{\rmn s} y_0^{(2)}\left(Az_0^{(2)}-C_{\rmn h}/C_{\rmn s}\right) \\
a_3 & = & C_{\rmn s} y_0^{(2)}\left[A \spg z_0^{(2)}\left(\nex+y_0^{(2)}\right)- \right.\nonumber \\ 
& -& \left. \left(A z_0^{(2)}-C_{\rmn h}/C_{\rmn s}\right)\left(G_{\rmn p}+\spg y_0^{(2)}\right)\right].
\eqe

\rmfamily
The number of real and complex roots of the equation $P(\lambda)=0$ can be determined, if one knows the 
signs of the constants. 
For values relevant to our physical problem, one finds that (i) either
all three constants are positive or (ii) only $a_2$ is negative. 
In the first case the polynomial has 3 negative real roots or 1 negative real root and two complex conjugates, while 
in the second case there is always 1 negative real root and 2 complex conjugates with positive real parts.

\begin{figure}
  \begin{tabular}{c}
\resizebox{\hsize}{!}{\includegraphics{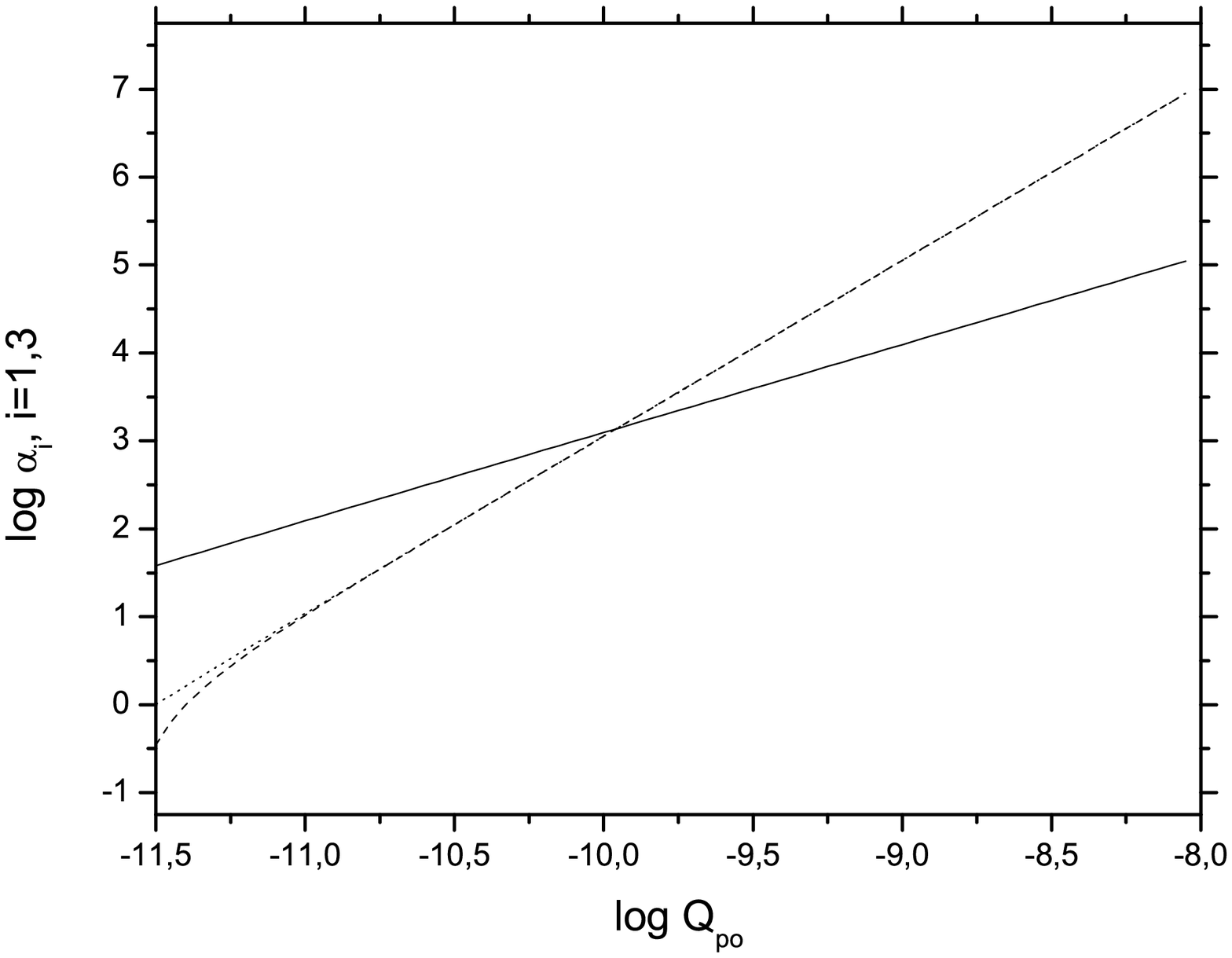}}\\
\resizebox{\hsize}{!}{\includegraphics{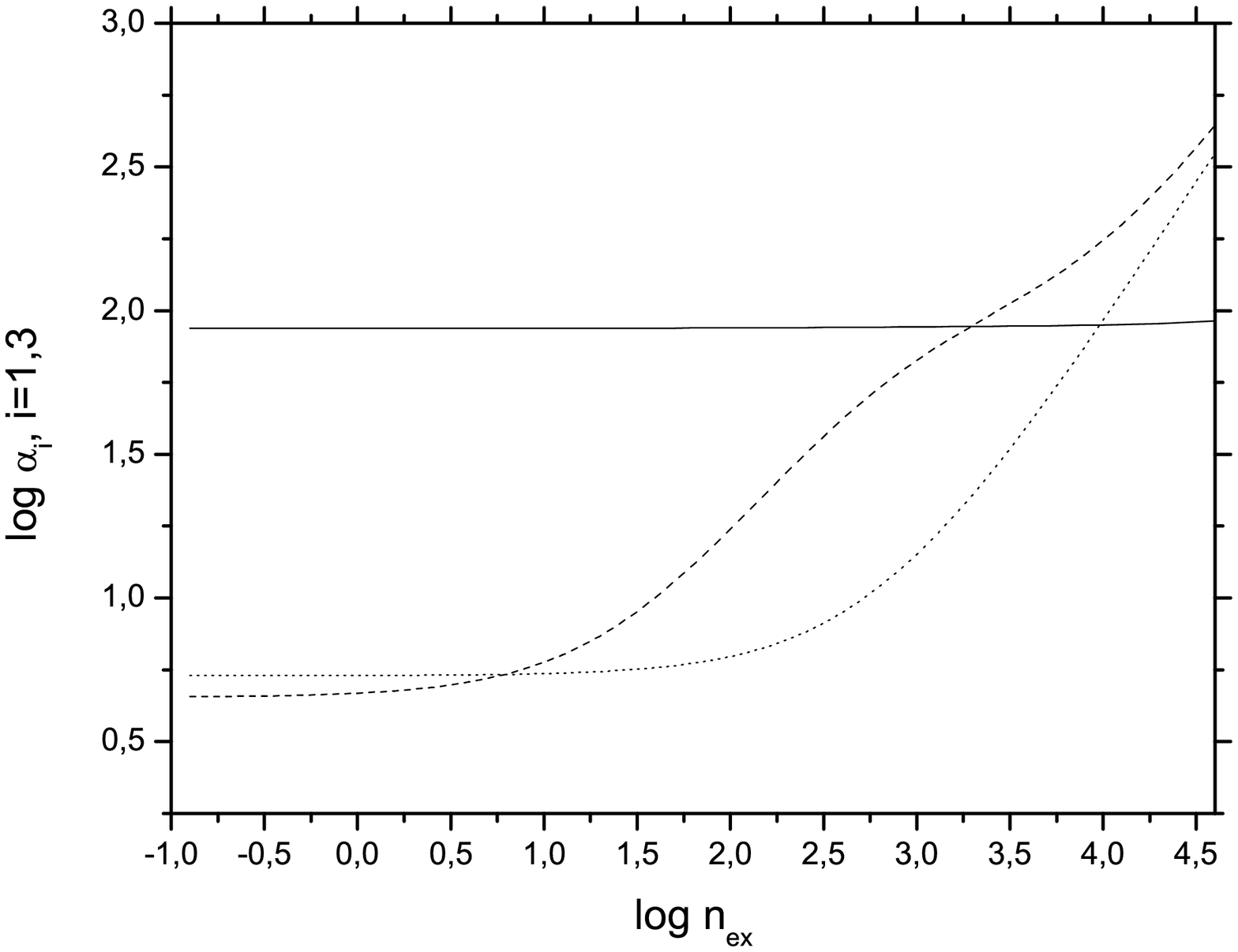}}
  \end{tabular}
\caption{Dependence of the constants of the characteristic polynomial $P(\lambda)$ on $\Qpo$ for $\nex=2$ (top panel) and
on $\nex$ for $\log \Qpo=-11.15$ (bottom panel), for a range of values relevant to the physical problem. In both panels solid, 
dashed and dotted lines represent
the constants $a_1$, $a_2$ and $a_3$ respectively. }
\label{poly}
\end{figure}

Figure \ref{poly} shows the dependence of constants $a_i$ on 
$\Qpo$ and $\nex$. Since a logarithmic scale is used, the negative values of $a_2$ are not shown.

\section[]{Derivation of the critical $\gamma$-ray compactness } 
Let us assume that hard photons are being injected into a spherical
source with a constant rate $Q_{\rmn h}^{\rmn {inj}}$ that corresponds to a compactness $\ell_{\rmn h}^{\rmn {inj}}$
 and that no soft photons are initially present
in the source. 
Automatic quenching of hard photons is possible if the injected compactness exceeds a certain value.
In this case, electron-positron pairs are being created spontaneously in the source, emitting synchrotron
radiation. Hard photons then undergo further absorption on the aforementioned soft photons and a non-linear
loop of processes begins to operate.
The equations that describe the above physical system can be written in the following form:
\eqb
\dot{n}_{\rmn h} & =& Q_{\rmn h}^{\rmn {inj}} -\nh -C_{\rmn h} \ns \nh \\
\dot{n}_{\rmn s} & = & -\ns +C_{\rmn s} \ns \nh,
\label{system-quench}
\eqe
where the constants $C_{\rmn h}$ and $C_{\rmn s}$ are defined in eq.~(\ref{constants1}).

There is a trivial stationary solution of the system (\ref{system-quench}):
$\bar{n}_{\rmn h}= Q_{\rmn h}^{\rmn {inj}}$, $\bar{n}_{\rmn s}=0$, that corresponds to the free propagation
of hard photons through the source, where pairs and soft photons are absent.
The Jacobian matrix evaluated for this solution has two real eigenvalues. 
For $Q_{\rmn h}^{\rmn {inj}}<1/C_{\rmn s}$
both are negative. Thus, the solution is stable. 
Moreover, in this case the system has no other physically
acceptable solution, i.e., $\ns>0$.
However, if $Q_{\rmn h}^{\rmn {inj}}>1/ C_{\rmn s}$ one of the eigenvalues becomes positive and the solution
with a zero soft photon population becomes unstable. Even a perturbation in the initially
absent soft photon distribution is sufficient for its subsequent growth. 
In this region, a second stationary solution of the system (\ref{system-quench})
appears. This is 
 $\bar{n}_{\rmn h}=1/C_{\rmn s}$,
$\bar{n}_{\rmn s}=\left(Q_{\rmn h}^{\rmn {inj}}-\bar{n}_{\rmn h} \right) / C_{\rmn h} \bar{n}_{\rmn h}$ and one can show that
for this region both eigenvalues of the corresponding Jacobian matrix are negative, i.e.,
the solution is stable.

Summarizing, the critical injection rate is $1/C_{\rmn s}$, which can be transformed to 
the critical compactness:
\eqb
\lhcr= \frac{\eh}{3C_{\rmn s}} \cdot  
\eqe

\bibliographystyle{mn2e}
\bibliography{1130.bib}

\label{lastpage}

\end{document}